\documentclass[aps, prc, twocolumn, tightenlines, letterpaper, preprintnumbers, floatfix, longbibliography, nofootinbib,superscriptaddress,10pt]{revtex4-2}

\usepackage[utf8]{inputenc}
\usepackage{amsmath}
\usepackage{graphicx}
\usepackage{dsfont}
\usepackage{physics}
\usepackage{placeins}
\usepackage{amssymb}
\usepackage{mathtools}
\usepackage{bbm}
\usepackage{amssymb}
\usepackage{tabularx}
\usepackage[colorlinks=true, allcolors=blue]{hyperref}
\usepackage{multirow}
\usepackage[T1]{fontenc}
\usepackage{bm}
\usepackage{xspace}
\usepackage[dvipsnames]{xcolor}
\usepackage[normalem]{ulem}
\usepackage{diagbox}
\usepackage{soul}
\usepackage{nccmath}
\usepackage{lipsum}
\usepackage{cancel}
\usepackage{bm}
\usepackage{algorithm2e}
\usepackage{tikz}
\usetikzlibrary{quantikz2}

\newcolumntype{Y}{>{\centering\arraybackslash}X}

\usepackage{orcidlink}

\definecolor{lavenderindigo}{rgb}{0.58, 0.34, 0.92}

\usepackage{stackengine}
\stackMath
\newcommand\tenq[2][1]{
 \def\useanchorwidth{T}%
  \ifnum#1>1%
    \stackunder[0pt]{\tenq[\numexpr#1-1\relax]{#2}}{\scriptscriptstyle\sim}%
  \else%
    \stackunder[1pt]{#2}{\scriptscriptstyle\sim}%
  \fi%
}

\makeatletter  
\renewcommand\onecolumngrid{
\do@columngrid{one}{\@ne}%
\def\set@footnotewidth{\onecolumngrid}
\def\footnoterule{\kern-6pt\hrule width 1.5in\kern6pt}%
}
\makeatother    

\begin{document}

\begin{figure}
\vskip -1.cm
\leftline{\includegraphics[width=0.15\textwidth]{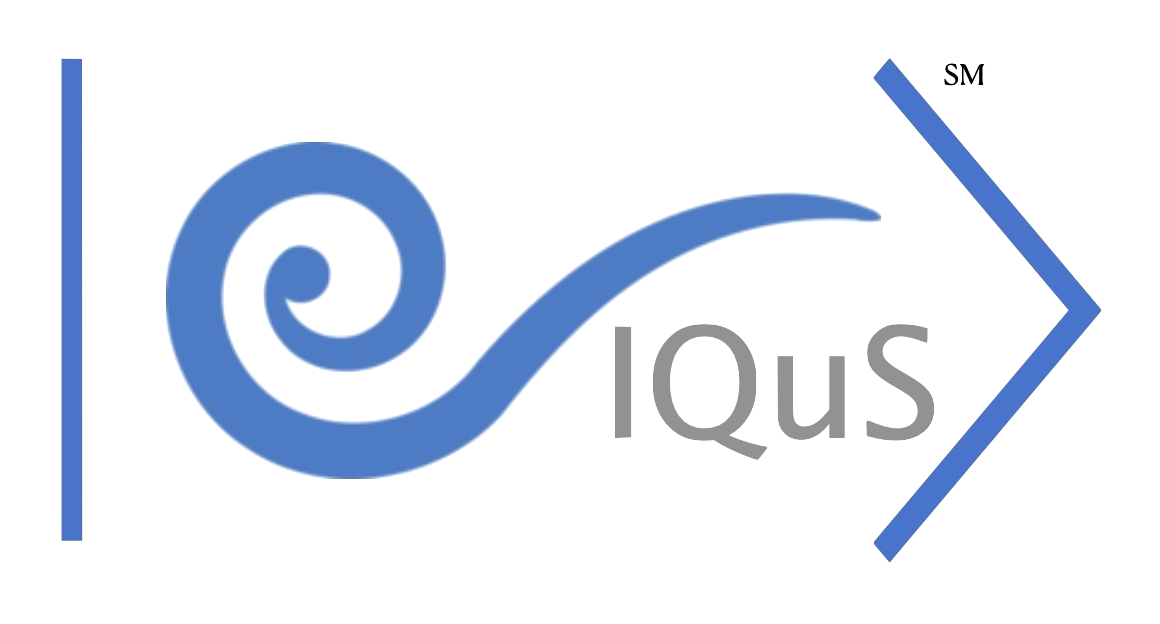}}
\end{figure}

\title{The Quantum Complexity of String Breaking in the Schwinger Model}

\author{Sebastian Grieninger\,\orcidlink{0000-0002-9523-5819}}
\email{segrie@uw.edu}
\affiliation{InQubator for Quantum Simulation (IQuS), Department of Physics, University of Washington, Seattle, WA 98195, USA}
\author{Martin J. Savage\,\orcidlink{0000-0001-6502-7106}}
\email{mjs5@uw.edu}
\affiliation{InQubator for Quantum Simulation (IQuS), Department of Physics, University of Washington, Seattle, WA 98195, USA}
\author{Nikita A. Zemlevskiy\,\orcidlink{0000-0002-0794-2389}}
\email{zemlni@uw.edu}
\affiliation{InQubator for Quantum Simulation (IQuS), Department of Physics, University of Washington, Seattle, WA 98195, USA}

\preprint{IQuS@UW-21-119, NT@UW-26-1}
\date{\today}

\begin{abstract}
\noindent 
String breaking, the process by which flux tubes fragment into hadronic states, is a hallmark of confinement in strongly-interacting quantum field theories.
A suite of quantum complexity measures is examined 
using Matrix Product States to characterize the string breaking process in the 1+1D Schwinger model.
We demonstrate the presence of nonlocal quantum correlations along the string that may affect fragmentation dynamics, and show that entanglement and magic offer complementary perspectives on string formation and breaking beyond conventional observables.
\end{abstract}
\maketitle

\noindent
{\it Introduction---}The formation of flux tubes, or chromoelectric strings, between color charges is a primary emergent feature of quantum chromodynamics (QCD)~\cite{Gross:1973id,Politzer:1973fx}.
At modest separations between static quark-antiquark pairs, the nonlinearity of the gluon self interactions~\cite{PhysRev.96.191}, combined with quantum fluctuations, confine the flux tubes into one-dimensional ``strings''.
As the separation increases, the energy stored in the flux tube grows approximately linearly until a two-hadron state becomes energetically favored.
Such final states correspond to dynamical quarks from the vacuum rearranging themselves into baryon number $\pm 1/3$ configurations around the static charges.
In nature, string breaking occurs in high-energy collisions that produce high-multiplicity final states of strongly-interacting particles.
While the dynamics of particle creation through string breaking, fragmentation and hadronization is well modeled~\cite{Field:1977fa,Andersson:1983ia,andersson1998lund,Sjostrand:2019zhc,GEANT4:2002zbu}, 
high-precision predictions in environments far from present-day experiments are challenging.
Such predictions will be crucial at the future Electron-Ion Collider for determining nucleon partonic distributions, measuring gluon helicity contributions, and understanding diffractive dijet production~\cite{Abir:2023fpo,AbdulKhalek:2021gbh}, 
and for the discovery of new fundamental physics at the LHC.

Significant theoretical and numerical results exist for string breaking processes 
(for recent overviews, see Refs.~\cite{Halimeh:2025vvp,Kharzeev:2026jkq}).
Classical lattice QCD simulations examine the electric potential between static charges as a function of their separation~\cite{PhysRevD.59.031501,CP-PACS:1998hwq,SESAM1998209,Bali:2005fu,Pennanen:2000yk,PhysRevD.63.111501,Bulava:2019iut}.
Hamiltonian and tensor network methods complement these techniques, providing access to real-time dynamics and quantum observables~\cite{Buyens:2015tea,Grieninger:2025rdi,Florio:2025hoc, Florio:2023dke, Florio:2024aix, Janik:2025bbz,Barata:2025hgx,Artiaco:2025qqq,Verdel:2019chj,Verdel:2023mmp,Mallick:2024slg} and allowing studies in two dimensions~\cite{Cochran:2024rwe,Gonzalez-Cuadra:2024xul, Borla:2025gfs,Cataldi:2025cyo,Xu:2025abo,DiMarcantonio:2025cmf,Ciavarella:2024fzw,Ciavarella:2024cyt}.
Recent advances in experimental quantum technologies have also enabled string breaking simulations on quantum hardware, e.g., Refs.~\cite{Crippa:2024hso,Liu:2024lut,De:2024smi,Surace:2024bht,Ciavarella:2024lsp,Alexandrou:2025vaj,Luo:2025qlg}.
While both static and dynamical systems have been explored, deep connections with confinement and screening remain to be established.

Measures of quantum complexity are sensitive probes of emergent phenomena~\cite{Haferkamp:2021uxo, Chitambar:2018rnj,Brown:2017jil,Eisert:2008ur,Leone:2021rzd,Robin:2020aeh,Haug:2023hcs,Tarabunga:2023hau,Hengstenberg:2023ryt,Haug:2024ptu,Emerson:2013zse,Howard:2017maw,Hamaguchi:2023zpb,Tirrito:2023fnw,Chernyshev:2024pqy,Cao:2024nrx,Robin:2024oqc,brokemeier2025quantum,Robin:2025ymq,Jiang:2025wpj}.
In this work, we investigate multipartite entanglement and magic in the ground state of the Schwinger model, a well-established testbed for quantum simulations of quantum field theories~\cite{PhysRevA.90.042305,Muschik:2016tws,Klco:2018kyo,Farrell:2023fgd,Farrell:2024fit,Nguyen:2021hyk}, as the system evolves from a confining string to isolated bound states (hadrons).
We describe minimal simulation requirements for lattice studies of string breaking, 
and perform simulations to study the inherent quantum complexity 
(quantum correlations) in the wavefunction, revealing previously unknown structure.
Measures of quantum complexity are found to change rapidly in the vicinity of 
string breaking,
offering a complementary view of mechanisms involved in the formation of hadrons and the associated vacuum rearrangement.


\noindent
{\it The Hamiltonian and Gauss's Law---}In 1+1D with open boundary conditions (OBCs), 
the distribution of fermion charges completely constrains the gauge field through Gauss's law.
Further, explicit gauge degrees of freedom are absent in axial gauge,
leaving a nonlocal fermionic interaction in the Hamiltonian induced by the Coulomb potential.
In the Kogut-Susskind formulation  \cite{Kogut:1974ag,Banks:1975gq}, with staggered fermion discretization \cite{Susskind:1976jm} and the Jordan-Wigner transformation to spin degrees of freedom~\cite{Jordan:1928wi,Lieb:1961fr}, the lattice Hamiltonian of the Schwinger model is (for a derivation, see Ref.~\cite{Banks:1975gq})
\begin{eqnarray}
	&& \hat{H} (d) \ =\   \frac{1}{4a}  \sum_{n=1}^{N-1}\left( \hat{X}_n \hat{X}_{n+1} + \hat{Y}_n \hat{Y}_{n+1} \right)
    \nonumber \\ 
    & &  +  \frac{m_\text{lat}}{2} \sum_{n=1}^{N} (-1)^n  \hat{Z}_n 
    +\frac{a}{2} \sum_{n=1}^{N-1} \left(\hat{E}_n +E_{\text{ext},n}(d)\right)^2
    \ ,
    \label{eq:h_obc}
\end{eqnarray}
where the lattice sites are labeled from $n=1$ to $N$, and the link indices range from $n=1$ to $N-1$.
Here, ${m_\text{lat}=m-\frac{g^2a}{8}}$~\cite{Dempsey:2022nys} is the chirally-improved (bare) fermion mass, $m$ and $g$ are the bare mass and coupling respectively and $N$ is the number of staggered sites (corresponding to 
$N_\text{phys}=N/2$
physical sites), and $a$ is the 
(staggered) lattice spacing.\footnote{We have not included constant terms in Eq.~\eqref{eq:h_obc}.
The physical lattice spacing is twice the staggered spacing, $a_{\rm phys}=2 a$,
and the number of physical lattice sites is half the number of staggered sites, $N_{\rm phys}=N/2$.
The physical length of the lattice is $L = aN = a_{\rm phys}N_{\rm phys}$.}
The last term specifies the nonlocal fermionic interaction, where 
\begin{align}
 \hat{E}_n \ =\  E_{0}+g \sum_{k=1}^{n} \hat{Q}_k \   , \quad 
 \hat{Q}_k \ = \ \frac{1}{2}\left((-1)^k+\hat{Z}_k\right)
 \ ,
 \label{eq:e_field_charge}
\end{align}
are the electric field between sites $n,n+1$, and the charge on site $k$, respectively.
By convention, we set the background field $E_0=0$.
A flux tube connecting two static external charges is represented by the external electric field $E_{\text{ext},n}(d)$~\cite{Grieninger:2025rdi}.\footnote{Instead of writing the electric field in terms of $\hat E_n$ and $E_\text{ext,n}(d)$ one could also modify Gauss's law as explained in~\cite{Buyens:2015tea}.}
A string of length $d$ centered on the middle of the lattice is implemented by applying the external field
\begin{equation}
	E_{\textrm{ext,}n}(d) \ = \ g\ \text{sign}(E)\cdot \Theta\!  \left( \frac{d/a - 1}{2} - \left| n - \frac{N}{2}\right|+\epsilon \right) \ ,
    \label{eq:L_ext}
\end{equation}
where $\Theta$ is the Heaviside step function and $\epsilon$ is a small number. 
This represents a string created by a static external fermion at $n=(N+1-d/a)/2$ and an antifermion at $n=(N+1+d/a)/2$. 

In the absence of dynamical screening and vacuum rearrangement, the ground state of the Hamiltonian~\eqref{eq:h_obc} encodes the rearrangement of the degrees of freedom in the vacuum in response to the insertion of static external charges.
We refer to the ground state of $H(d=0)$ as the ``vacuum'', 
and study the behavior of the ground states of $H(d)$ as a function of $d$.

\noindent
{\it Minimal lattice requirements---}In lattice simulations, physical states must be well-contained within the volume of the lattice, so that boundary effects can be quantified and systematically removed. 
The Compton wavelength $l$ of excited states relevant to the physics of the process must be much smaller than the lattice size $L$. 
Furthermore, the resolution of low-lying state wavefunctions must be fine enough to resolve quantities of interest perturbatively close to their continuum values, meaning that $l/a_{\rm phys}\gg1$.
Therefore, the lattice parameters must satisfy 
$\frac{1}{L} \ll \Delta \ll \frac{1}{a_{\rm phys}}$ 
for a lattice simulation to reliably capture the physics of the process. 
Here $\Delta \sim 1/l$ is the gap to the first excited state.\footnote{As an example, for lattice QCD calculations at the physical values of the quark masses, 
with $m_\pi\sim 140~{\rm MeV}$,
this condition implies that $m_\pi L\gg 1$ and $m_\pi a\ll 1$. 
With a lattice spacing of $a=0.1~{\rm fm}$, this gives $m_\pi a\sim 0.07 \ll 1$, 
and 64 lattice sites in each spatial direction gives $m_\pi L\sim 4.5 \gg 1$.
}
When simulating confining systems with static background charges separated by a distance $d$ (or multiparticle states), boundary effects on the perturbations must be exponentially suppressed.
This requires $L\gg d$.\footnote{For a simple system of charges on a 1D lattice, the potential experienced by a static charge is the sum over image charge contributions~\cite{Huang:1957im,Hamber:1983vu,Luscher:1986pf,Luscher:1990ux,Luscher:1990ck,Detmold:2007wk,Lu:2018pjk}. 
For PBCs, ${V^{\rm eff}(d)  =  \sum_{n=-\infty}^{+\infty}\  V(|d+nL|) }$.
Twisted boundary conditions can be helpful in minimizing the effects of the boundary, see e.g., Ref.~\cite{Briceno:2013hya}, particularly i-PBCs for which the leading contributions to single-particle observables vanish.
Boundary effects are discussed in App.~\ref{app:obc_vs_pbc}.
}

In this work, we select parameters for the spectrum and interactions to satisfy these conditions.
As a starting point, we chose 
$g=0.09$, 
$m_\text{lat}=0.045$, $a=1$ and $N=220$ staggered lattice sites ($L=110$)
in Eq.~(\ref{eq:h_obc}).

With these parameters, the vector meson mass is found to be $M_v=0.13647$, so that $M_v a_{\rm phys} \sim 0.2729$ and $M_v L \sim 15.0117$, 
satisfying the conditions for perturbative finite volume and lattice spacing corrections.
Further, 
the finite volume corrections are suppressed by 
${\cal O}(e^{-M_v L/2}) \sim 2\times 10^{-4}$ for $d=L/2$.
Continuum physics is identified from the $a\rightarrow 0$, $N\rightarrow\infty$
limit of the Hamiltonian in
Eq.~(\ref{eq:h_obc}) with fixed $m$ and $g$.
To approach the continuum with this physical parameter set, we perform simulations with 
$\{N,a\} = \{220,1\}, \{440,\frac{1}{2}\}$ and $\{880,\frac{1}{4}\}$ so that $L=110$.

\noindent
{\it Measures of quantum complexity---}The distribution of quantum information in a state provides insight into its underlying structure.
In gauge theories, physically-meaningful measures of entanglement and quantum complexity must be gauge invariant, which restricts their definition to operations that preserve global charges.
The ground state wavefunction $|\psi\rangle$  of the Schwinger model has vanishing total electric charge, $Q=0$.
However, reduced states have contributions from all charge sectors.
For a bipartition of the lattice into regions $A$ and $B$, the reduced density matrix $\hat{\rho}_A$ has a block diagonal structure, with the blocks characterized by the charge within region $A$,
\begin{eqnarray}
    |\psi\rangle & \ = \ & \sum_{Q,i}\ c^{(Q)}_i |\psi^{(Q,i)}_A\rangle \otimes |\psi^{(-Q,i)}_B\rangle
    \ \ , \nonumber\\ 
    \hat{\rho}_A & \ =\  &   {\rm Tr}_B \left[ \hat{\rho}_{AB} \right]\ =\ \sum_Q \ p_Q \ \hat{\rho}^{(Q)}_A
    \ ,
\label{eq:ABQ}
\end{eqnarray}
where 
$\hat{\rho}_{AB}=|\psi\rangle\langle\psi|$.
The block-diagonal form of $\hat{\rho}_A$ allows the entanglement entropy to be decomposed 
into contributions from the charge sectors~\cite{Ghosh:2015iwa,Turkeshi:2020yxd,Buividovich:2008gq,Donnelly:2011hn,Casini:2013rba,Radicevic:2014kqa,Aoki:2015bsa,Soni:2016ogt,Goldstein:2017bua,Nishioka:2018khk,Amorosso:2024leg,Amorosso:2024glf}
\begin{eqnarray}
S(\hat{\rho}_A) & \  = \ & 
\sum_Q p_Q \ S(\hat{\rho}^{(Q)}_A) - \sum_Q p_Q \log_2 p_Q
    \ ,
\label{eq:SEQ}
\end{eqnarray}
where $S(\hat{\rho})=-{\rm Tr} \left[ \hat{\rho}\log_2\hat{\rho} \right]$ is the von Neumann entanglement entropy.
By treating each charge sector separately, this quantity is naturally gauge invariant.
This definition extends to other quantities, including mutual information  (MI) and antiflatness.
The MI is defined by
\begin{eqnarray}
I(A:B) & \ = \ & S(\hat{\rho}_A) + S(\hat{\rho}_B) - S(\hat{\rho}_{AB})
    \ ,
\label{eq:IAB}
\end{eqnarray}
using the definition in Eq.~(\ref{eq:SEQ}).
The antiflatness for the same bipartition, 
$ {\cal F}(\hat{\rho}_A)$, 
which provides a lower bound to the nonlocal magic~\cite{Cao:2024nrx},\footnote{
Results from small-qubit systems suggest the nonlocal magic is related to its lower bound by ${\cal M}_2^{(\text{NL})}=4{\cal F}$.} becomes 
\begin{eqnarray}
{\cal F}(\hat{\rho}_A) & \ = \ & \sum_Q p_Q^2\  {\cal F}_A(\hat{\rho}_A^{(Q)})
\ ,
\nonumber\\
{\cal F}_A(\hat{\rho}) & \ = \ &  {\rm Tr}\hat{\rho}^3 -\ \left({\rm Tr}\hat{\rho}^2\right)^2 \ =\ {\rm Var}(\hat{\rho}^2)
    \ .
\label{eq:AFQ}
\end{eqnarray}

The recently identified Stabilizer Renyi Entropies (SREs) 
are calculable measures of the quantum magic 
in both pure and mixed states but they are not monotones for mixed states.
The density matrix of a pure state $|\psi\rangle$ may be written in the Pauli basis,
\begin{equation}
    \hat{\rho} \ = \ \ket{\psi} \bra{\psi} \ = \ 
    \frac{1}{\mathbf d} \sum_{\hat P } \langle \psi |\hat{P} | \psi \rangle \, \hat{P} 
    \ =\ 
    \frac{1}{\mathbf d} \sum_{\hat P } c_P \, \hat{P} 
    \; ,
\end{equation}
where $c_{\hat{P}} \equiv \langle \psi |\hat{P} | \psi \rangle$,
and ${\mathbf d}=2^{n_Q}$ for $n_Q$  qubits.
For pure states,
the quantity $\Xi_{\hat{P}} \equiv   c_{\hat{P}}^2/{\mathbf d}$
is a probability distribution~\cite{Leone:2021rzd}.
For a stabilizer state $\ket{\psi}$,
$c_{\hat{P}} = \pm 1$ for ${\mathbf d}$ commuting Pauli strings
and $c_{\hat{P}} = 0$ for the remaining  ${\mathbf d}^2-{\mathbf d}$ strings~\cite{zhu2016clifford}.
The SREs measure the deviation from stabilizer states, 
\begin{equation} 
\mathcal{M}_{\alpha}(\ket{\psi}) \ = \ -\log_2 {\mathbf d} + \frac{1}{1-\alpha} \log_2 
\left( \sum_{\hat{P} } \Xi_{\hat{P}}^{\alpha} \right) \; ,
\label{eq:Renyi_entropy_def1}
\end{equation}
with ${\cal M}_2$ used in this work:
\begin{eqnarray}
    {\cal M}_2 & \ = \ &    -
    \log_2 \ {\mathbf d} \sum_{\hat{P} }  \Xi_{\hat{P}}^2
\ .
    \label{eq:MlinM1M2_def}
\end{eqnarray}

Nonlocal magic of a bipartition of a system into regions A and B, is the amount of magic that cannot be removed by local unitary transformations.
Operationally, it is defined by 
\begin{eqnarray} 
\mathcal{M}^{(\text{NL})}_{\alpha}(\ket{\psi}) & \ = \ & 
\min_{ {\bm\theta}_A,  {\bm\theta}_B} 
\mathcal{M}_{\alpha}(U_A({\bm\theta}_A) U_B({\bm\theta}_B)\ket{\psi}) 
\ ,
\label{eq:NLMAGIC}
\end{eqnarray}
for charge-preserving transformations to maintain gauge invariance.
An upper bound for the nonlocal magic ${\cal M}_2$ is found by writing the 
bipartitioned $Q=0$ ground state wavefunction as
in Eq.~(\ref{eq:ABQ})~\cite{Falcao:2024msg},
 \begin{eqnarray}
&& e^{-{\cal M}_2}\ =\   
 \sum_{ {\bm\sigma}^{(1)} , {\bm\sigma}^{(2)} , {\bm\sigma}^{(3)} , {\bm\sigma}^{(4)}}
 c_{{\bm\sigma}^{(1)}}\ 
 c_{{\bm\sigma}^{(2)}}\ 
 c_{{\bm\sigma}^{(3)}}\ 
 c_{   {\bm\sigma}^{(1)}   {\bm\sigma}^{(2)}   {\bm\sigma}^{(3)}   }\ 
 \nonumber\\
 &&
 \qquad\times
 c^*_{   {\bm\sigma}^{(1)}   {\bm\sigma}^{(2)}   {\bm\sigma}^{(4)}   }\ 
 c^*_{   {\bm\sigma}^{(1)}   {\bm\sigma}^{(3)}   {\bm\sigma}^{(4)}   }\ 
 c^*_{   {\bm\sigma}^{(2)}   {\bm\sigma}^{(3)}   {\bm\sigma}^{(4)}   }\ 
 c^*_{{\bm\sigma}^{(4)}}
 \ ,
 \label{eq:Tara}
 \end{eqnarray}
where 
$c_i= c^{(Q)}_i$
and products of ${\bm\sigma}^{(s)} $ denote elementwise multiplication. 

The Robustness of Magic  (RoM)~\cite{Howard:2017maw,Pashayan:2015cos,Heinrich:2019aei} is defined by the minimum distance to the surface of stabilizer states,
\begin{eqnarray}
    R(\hat{\rho})  & \ = \ &  \min_{\bm x} \left\{
    \ ||\bm x||_1 \ \ \  \bigg\rvert \ \ \ \hat{\rho} = \sum_{i}
     x_i\hat{\rho}_{s_i}\right\} 
     \ ,
\label{eq:RoMdef}
\end{eqnarray}
given by the 1-norm of the coefficients of stabilizer density matrices $\hat{\rho}_{s_i}$.
States that are elements of the stabilizer polytope will have $x_i>0$ and $R(\hat{\rho})=1$, while nonstabilizer states will have one or more $x_i<0$ and thus $R(\hat{\rho})>1$.
As the trace of the density matrix is unity, $R(\hat{\rho})$ measures the amount of negativity in the 
expansion coefficients,
\begin{eqnarray}
||\bm x||_1 \ = \ 1 + 2 \sum\limits_{i, x_i<0} |x_i|
\ ,
\end{eqnarray}
and hence provides a measure of the difficulty for classical simulation
(more specifically, $R(\hat{\rho})-1$).
This is a faithful magic monotone for both pure and mixed states.
The nonlocal RoM (NL RoM) associated with subsystems $A$ and $B$ is defined by a further minimization with respect to charge-preserving 
local unitary transformations in each region,
\begin{eqnarray}
    && R_{AB}^{({\rm NL})}(\hat{\rho}_{AB})  \ =\   \min_{{\bm x}, {\bm\theta}_A,  {\bm\theta}_B} 
    \left\{
    \ ||{\bm x}||_1 \ \ \  \bigg\rvert \ \ \ 
    \right.\nonumber\\
    & & \left.
     \hat{U}_A({\bm\theta}_A) \hat{U}_B({\bm\theta}_B) \hat{\rho}_{AB} \hat{U}_A^\dagger ({\bm\theta}_A) \hat{U}_B^\dagger ({\bm\theta}_B)  = \sum_{i}
     x_i\hat{\rho}_{s_i}
     \right\} 
     \ .
\label{eq:NLRoMdef}
\end{eqnarray}
The transformations are constrained to preserve charge in each region
for a gauge-invariant definition of $R_{AB}^{({\rm NL})}$.
In the Schwinger model with the mapping to qubits 
as in Eq.~\eqref{eq:h_obc}, 
for regions $A$ and $B$ each having support on two qubits, the transformations are each parameterized by four angles 
\begin{eqnarray}
    \hat{U}({\bm \theta}) & \ = \ & e^{-i \left( \theta_0 \hat{Z}\hat{I} + \theta_1 \hat{I}\hat{Z} + \theta_2\hat{Z}\hat{Z} + \theta_3 (\hat{X}\hat{X}+\hat{Y}\hat{Y}) \right)}
    \ \ ,
\label{eq:QpresTrans}
\end{eqnarray}
as opposed to the 15 angles for an arbitrary SU(4) transformation.

\begin{figure}
     \centering
     \includegraphics[width=\linewidth]{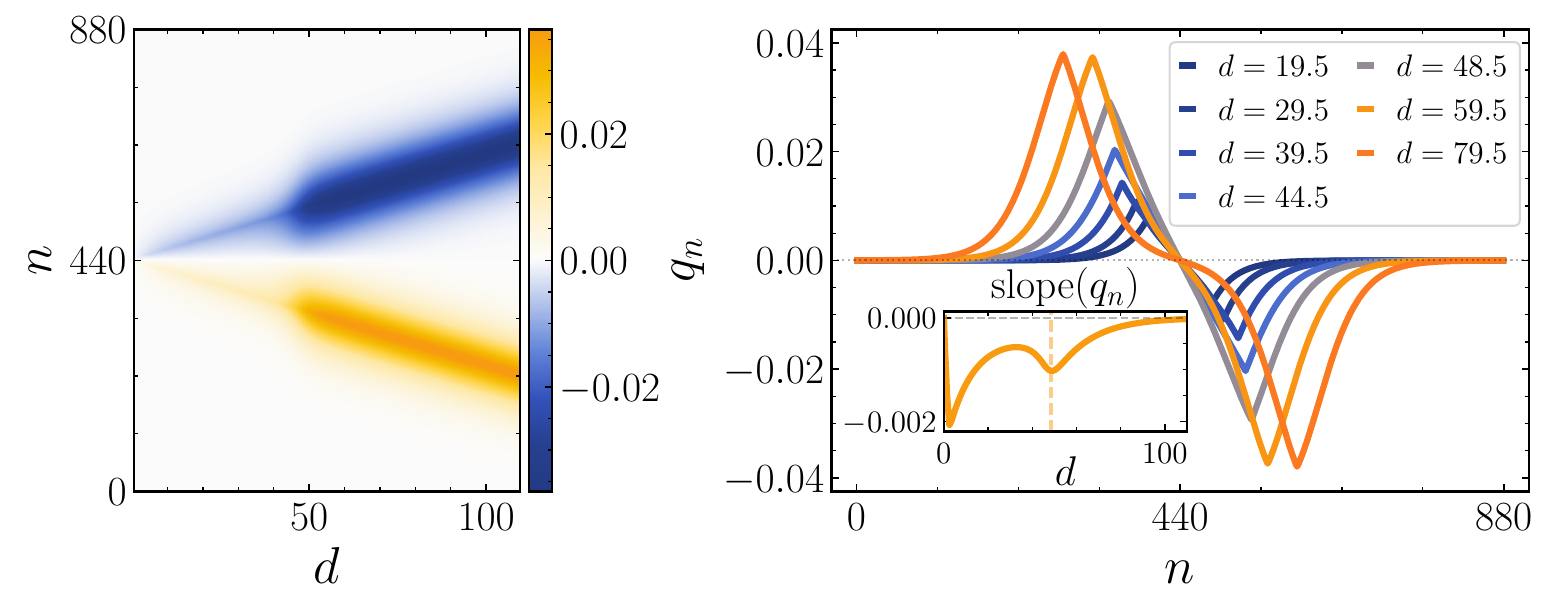} 
     \caption{{\it The charge density $q_n$ through the string breaking process.}
     Left: the vacuum-subtracted $q_n$ obtained with simulation parameters described in the text using $N=880$ and $a=1/4$, as a function of position $n$ and external charge separation $d$. 
     Right: cross sections of $q_n$ for a selection of $d$'s. 
     The inset shows the slope of $q_n$ at the center of the lattice, which has a local extremum at $d=48.5$.
     }
     \label{fig:chargespatial}
\end{figure}

\noindent 
{\it Results---}We first show results obtained for classical local observables, conventionally studied in the context of string breaking.
The left panel of Fig.~\ref{fig:chargespatial} shows the vacuum-subtracted charge density $q_n = \langle \hat{Q}_n\rangle/a$ as a function of the static charge separation $d$. 
With the simulation parameters described in the previous section, a small range of $d$ captures significant changes in the structure of the ground state wavefunction. 
Dramatic changes in the charge distribution around $d=45-50$ reflect the transition from a single quarkonium-type state with a string between the static charges to a molecular-type state~\cite{Guo:2017jvc} of two mesons. 
As the external charges separate beyond this point, the mesons become isolated and the string between them vanishes.

The charge distribution between the static charges 
(right panel of Fig.~\ref{fig:chargespatial}) 
is observed to transition from linear to curved with increasing $d$,
consistent with charge localization due to confinement.
The inset shows the profile of the charge distribution.
The width of this peak provides a measure of the separations over which the ground state structure changes from string-dominated to meson-dominated.
The accompanying energy-momentum tensor components, 
chiral condensate and electric field 
are discussed in App.~\ref{app:classical}. 
These classical quantities have been well-studied in the context of string breaking (e.g.,~\cite{Buyens:2015tea,Grieninger:2025rdi}) and our results agree with existing literature.

Quantum correlations across a bipartition of the system at the string center constitute an informative class of complexity measures.
The entanglement entropy of such bipartitions has been studied previously (e.g., Refs.~\cite{Grieninger:2025rdi,Florio:2025hoc}).
For this half-lattice bipartition, 
we examine the (gauge-invariant) entanglement entropy using Eq.~\eqref{eq:SEQ}, the lower bound of the nonlocal magic using the antiflatness (Eq.~\eqref{eq:AFQ})
and the upper bound of the nonlocal magic using Eq.~\eqref{eq:Tara}.
\begin{figure}
     \centering
     \includegraphics[width=\linewidth]{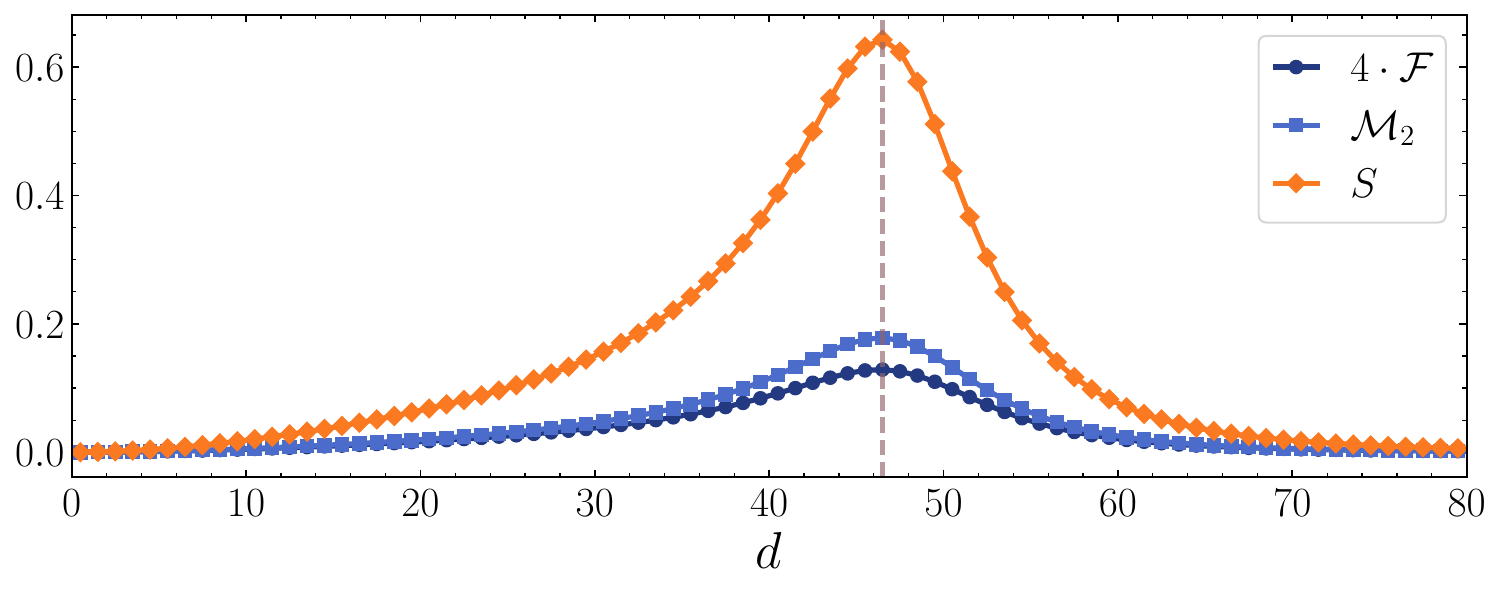} 
     \caption{{\it Bipartite measures of entanglement and quantum complexity in string breaking.}
     The symmetric bipartition entanglement entropy, antiflatness and the upper bound on the nonlocal ${\cal M}_2$ using simulation parameters $N=880, a=1/4, g=0.09, m=0.04601$, as a function of separation between static sources in terms of physical spatial sites.
     These quantities all peak at $d=46.5$ (see Table~\ref{tab:peaks}).
     }
     \label{fig:EEandAF}
\end{figure}
Figure~\ref{fig:EEandAF} shows these quantities as a function of $d$.  
The previously identified entanglement entropy peak in the vicinity of string breaking is also observed in both the lower and upper bounds of nonlocal magic.
This result is consistent with the nonlocal magic being driven by entanglement between the regions.\footnote{
The bounds provide all the accessible information about bipartition nonlocal magic because the required minimizations are impractical for the lattice sizes considered in this work.}

We conclude that the nonlocal magic increases during string breaking before returning to its vacuum value as the mesons separate.
This is typical of a transition from long-range to short-range correlations.
Moreover, this is consistent with the bipartition surface being (exponentially) insensitive to the presence of one or more mesons located in the bulk of the lattice, far from boundaries.
This can be considered a magic barrier, analogous to those observed in the time evolution of nonequilibrium systems~\cite{Ebner:2025pdm}.

We probe quasi-local quantum correlations with the $n$-tangle $\tau^{(n)}$~\cite{PhysRevA.63.044301}
\begin{eqnarray}
    \tau^{(n)}_{(i_1 ... i_n)} \ & = & \ |\langle \psi | 
    \hat Y_{i_1}\otimes \dots \otimes \hat Y_{i_n}
    | \psi^* \rangle|^2 \ ,
\label{eq:n-tangle}
\end{eqnarray}
where $\hat{Y}_{i_k}$ is the Pauli matrix acting on qubit $i_k$. 
Specifically, the 2-tangle $\tau^{(2)}_{(i,i+1)}$ provides a measure of nearest-neighbor entanglement.
\begin{figure}
     \centering
   \includegraphics[width=\linewidth]{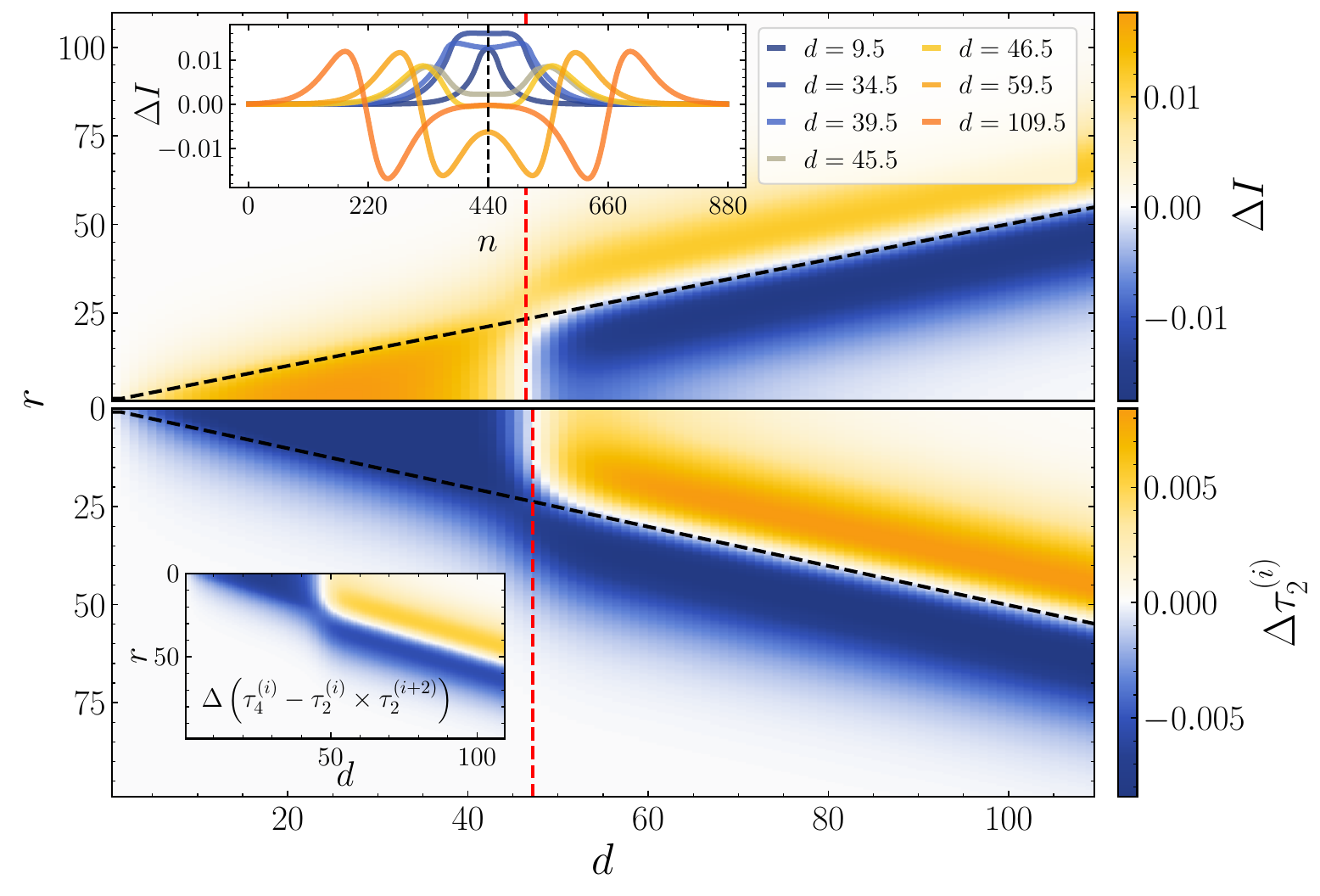}  
     \caption{{\it Local and multipartite entanglement in string breaking.} 
     Top: the local vacuum-subtracted MI as a function of distance from the center $r$ and external charge separation $d$ for $N=880, a=1/4, g=0.09, m=0.04601$.
     The inset shows cross sections of the MI for a selection of $d$'s.
     Bottom: the vacuum-subtracted 2-tangle for the same parameters. 
     The inset on the bottom panel shows the vacuum-subtracted 4-tangle. The black dashed lines indicate the peak of the charge distribution shown in Fig.~\ref{fig:chargespatial}. The red dashed lines show $d$ where the respective quantity changes sign for the smallest $r$.
}
     \label{fig:tau2}
\end{figure}
The lower panel of Fig.~\ref{fig:tau2} shows the vacuum-subtracted $\tau^{(2)}_{(i,i+1)}$ 
across the lattice as a function of $d$.
The nearest-neighbor entanglement is modified along the length of the string, which rapidly vanishes, changes sign in the vicinity of string breaking, then becomes confined into the mesons as they become isolated.
This behavior is also evident in the vacuum-subtracted 4-tangle, 
$\tau^{(4)}_{(i,i+1,i+2,i+3)} - \tau^{(2)}_{(i,i+1)} \tau^{(2)}_{(i+2,i+3)}$, 
(inset of lower panel in Fig.~\ref{fig:tau2})
and in the vacuum-subtracted MI 
$\Delta I$,
(upper panel of Fig.~\ref{fig:tau2}),
both between adjacent spatial sites. 
The shape looks similar to $\tau^{(2)}_{(i,i+1)}$,
indicating that much of the information is contained in the 2-tangle.\footnote{
We find the quantity 
${\tau^{(4)}_{(i,i+1,i+2,i+3)} -c\, \tau^{(2)}_{(i,i+1)} \tau^{(2)}_{(i+2,i+3)}\rightarrow 0}$ 
for $c\approx 1.78$. 
Similarly, the nonlocal 4-tangle ${\tau^{(4)}_{(N/2,N/2+1,i,i+1)} - \tau^{(2)}_{(N/2,N/2+1)}\tau^{(2)}_{(i,i+1)} }\to 0$ for $i$ away from the center.}
This suggests that (local) entanglement structure is limited to physical sites.
There is a separation for which $\tau^{(2)}_{(i,i+1)}$ vanishes along much of the string (indicated by the red dashed line), coinciding with the point of string breaking.
Notably, the zero that develops in $\tau^{(2)}$ and $\Delta I$ tracks the maximum of the charge distribution (shown with the black dashed line) after the string breaking. 
Deviations of $\Delta I$ from 0 show the localization of correlations as a function of $d$, reflecting the underlying structure of string breaking and meson formation.
The $\Delta I$ in the central region after string breaking arises from the nuclear force between the mesons, which vanishes when the mesons are sufficiently separated. 
It is interesting to note, that both the maximum and minimum of $\Delta I$ are within the width of the meson defined by the charge profile peak (Fig.~\ref{fig:chargespatial}).

We use the RoM, defined in Eq.~(\ref{eq:RoMdef}), as a measure of both local and nonlocal magic.
\begin{figure}
     \centering
     \includegraphics[width=\linewidth]{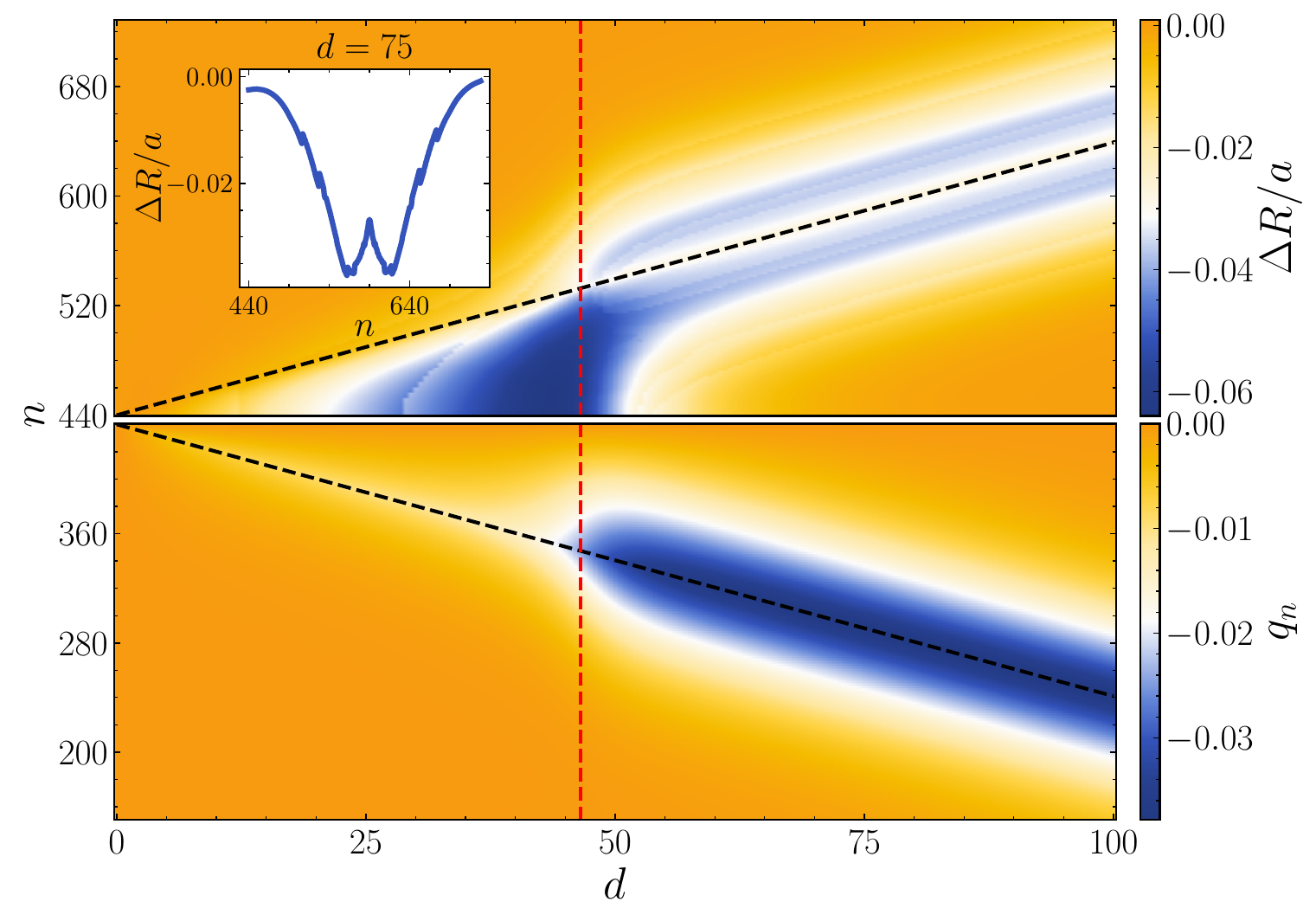}
     \caption{{\it Internal structure of the outgoing meson revealed by the RoM.} 
     Top: the vacuum-subtracted RoM of adjacent sites, $\Delta R$, in units of $a$, shown
      for $n \geq 440$ as a function of $d$, for $N=880,a=1/4,g=0.09, m=0.04601$.
     The inset shows $\Delta R$ inside the meson region for $d=75$.
     Bottom: the charge density $q_n$ for $n\leq440$ for the same system.
     The black dashed lines show the external charge positions, and the red dashed lines mark the peaks of the bipartite complexity measures shown in Fig.~\ref{fig:EEandAF}.}
     \label{fig:ROMadj}
\end{figure}
Figure~\ref{fig:ROMadj} shows the vacuum-subtracted RoM,
$\Delta R$, defined via Eq.~(\ref{eq:RoMdef}) as
\begin{eqnarray}
    \Delta R(i) \ & = & \ R(i:i+3) - R_{\Omega}(i:i+3)
    \ ,
    \label{eq:R4}
\end{eqnarray}
of four adjacent lattice sites, $i, \dots,  i+3$,
as a function of position and the separation of the external charges.
Here $R_{\Omega}$ denotes the value at $d=0$.
With increasing $d$, the deviation of $\Delta R$ from zero increases along the string, 
becoming maximal at the point where the string breaks, then decreasing rapidly as the mesons separate.
The (local) RoM reveals structure in the string state before it breaks, and in the outgoing mesons.

The inset of Fig.~\ref{fig:ROMadj} shows $\Delta R$ within the meson at $d=75$. 
This suggests that moments of RoM distributions in hadronic structure calculations could be complementary probes of structure, in the same way that moments of the charge or axial distributions are computed (and measured experimentally). 
This shares similarities with the behavior of $\Delta I$ within the meson (inset of top panel in Fig.~\ref{fig:tau2}).
The dependence of $\Delta R$ on lattice spacing and $m/g$ is discussed in App.~\ref{app:supplemental_results}.

A novel class of quantum correlations in the context of string breaking are those between 
spatially-separated subregions~\cite{Qian:2025oit}.
The mutual information $\Delta I(A:B)$ provides a measure of both classical and quantum correlations between regions $A$ and $B$.
\begin{figure}
     \centering
     \includegraphics[width=\linewidth]{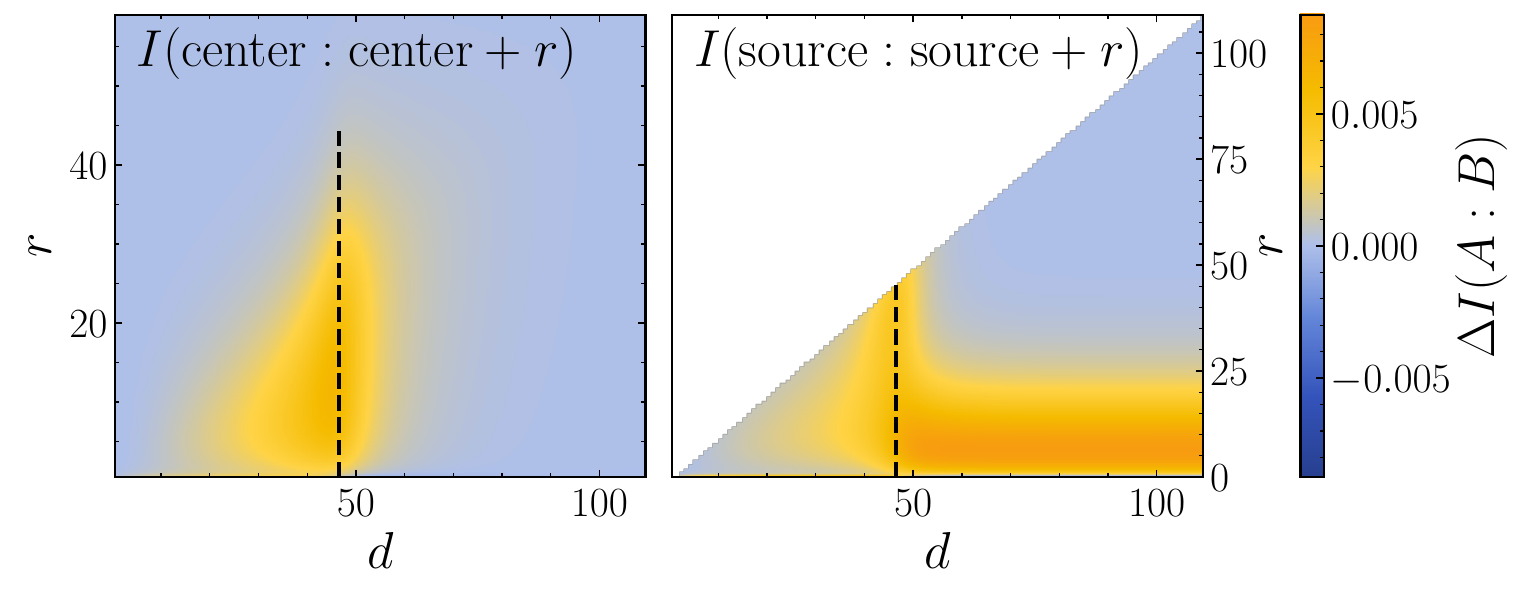} 
     \caption{{\it Mutual information in string breaking.} 
     Left: The vacuum-subtracted MI,
      $\Delta I(A:B)$, where region $A$ is at the center of the lattice, $A=\left(\frac{N}{2}+1,\frac{N}{2}+2\right)$ and $B$ is a distance $r$ away, $B=A+(r,r+1)$.
      The parameters $N=880,a=1/4,g=0.09, m=0.04601$ are used.
      Right: $\Delta I(A:B)$ where $A$ is at the left external charge, $A=\left(\frac{N}{2}-\frac{d}{2a}-1,\frac{N}{2}-\frac{d}{2a}\right)$, and $B$ is a distance $r$ away, $B=A+(r,r+1)$.
      The dashed lines mark the peak of the bipartite complexity measures shown in Fig.~\ref{fig:EEandAF}.}
\label{fig:MI}
\end{figure}
The left panel of Fig.~\ref{fig:MI} shows  $\Delta I(A:B)$ between spatial sites along the string with respect to the center of the system.
The MI 
is continuous along the string, peaking at the point of string breaking, and rapidly returning to  vacuum values thereafter.
Perhaps it is not surprising that the string supports long-distance correlations (classical or quantum) along the length of the string, which vanish as the system transitions to isolated mesons.
However, the transition region extends (at least)
over the range of the nuclear force between the mesons.\footnote{
The behavior of $\Delta I$ and $\Delta R^{\text{(NL)}}_{AB}$ for a system with unnatural scattering parameters may determine whether the string-breaking scale, the nuclear force or the scattering parameters drive changes in quantum complexity.}
The right panel of Fig.~\ref{fig:MI} shows $\Delta I$ over the interval between the static charges, where region $A$ is fixed at the position of the left charge. 
Its behavior along the string is analogous to the left panel, and it is nonzero over the extent of the meson for larger $d$.

\begin{figure}
     \centering
     \includegraphics[width=\linewidth]{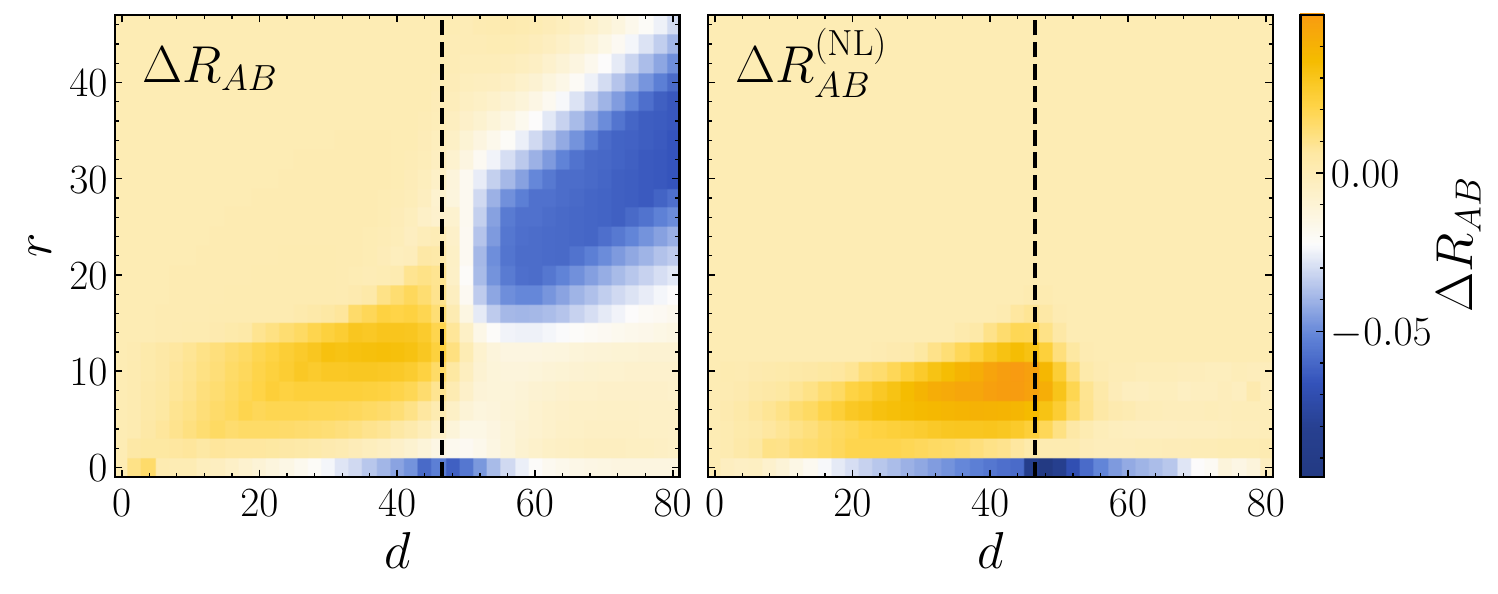}
     \caption{{\it Nonlocal quantum correlations in string breaking as measured by the NL RoM.}
     Left: the RoM of disjoint regions, $\Delta R_{AB}$, where region $A$ is at the center of the lattice and region $B$ is a distance $r$ away. 
     The vacuum value ($d=0$) and the values as $r\to\infty$ are subtracted.
     The system parameters $N=220,a=1,m_\text{lat}=0.045,g=0.09$ are used.
     Right: the NL RoM, $\Delta R^\text{(NL)}_{AB}$ for the same system parameters.
     The dashed lines mark the peak of the bipartite complexity measures shown in Fig.~\ref{fig:EEandAF}.
     }
     \label{fig:RoM}
\end{figure}
The left panel of Fig.~\ref{fig:RoM} shows the RoM $\Delta R_{AB}$ between disjoint subsystems on the lattice as a function of the distance between the regions $r$.
As in the left panel of Fig.~\ref{fig:MI}, region $A$ is at the center of the lattice, ${A=(N/2+1,N/2+2)}$, and region $B$ is distance $r$ away, 
${B=A+(r,r+1)}$.
We define this quantity with a different vacuum subtraction to that in Eq.~(\ref{eq:R4}),
\begin{eqnarray}
    \Delta R_{AB} \ & = & \ R(A,B) - R_{\infty}(A,B) - R_{\Omega}(A,B)
    \ ,
    \label{eq:R4AB}
\end{eqnarray}
where $R_{\infty}$ 
denotes the quantity evaluated as $r\rightarrow\infty$ 
and $R_{\Omega}$ denotes the value at $d=0$ (as before).
The reason for this is that the wavefunction is modified at the center of the lattice 
before string breaking, providing a modification in $R$ for all values of $r$.
Unlike $\Delta I$, $\Delta R_{AB}$ exhibits a localized rapid change of sign at the point of string breaking.
This is consistent with the center being the location of the most ``restructuring'' of the wavefunction.
This increase in charge density seen in Fig.~\ref{fig:chargespatial} 
coincides with the ``hot spot'' in the left panel of Fig.~\ref{fig:RoM}.
The behavior of $\Delta R_{AB}$ in the transition region immediately after string breaking may reveal new insights into the strong nuclear force.
Interestingly, the NL RoM (defined in Eq.~\eqref{eq:NLRoMdef}) shows similar, but more localized behavior to the MI (cf. Fig.~\ref{fig:MI}). 
The outgoing meson ``track'' vanishes in the NL components of the RoM (right panel of Fig.~\ref{fig:RoM}), suggesting that  NL quantum information is only encoded in the string state.
The presence of NL RoM establishes a link between spatially extended physical objects (strings) and nonlocal quantum complexity independent of basis choice.
These purely quantum correlations, 
absent from traditional models of hadronization, 
could imprint themselves into final states in high-energy collisions.

\noindent
{\it Discussion---} 
Continuing along the path of scientific discovery in high-energy and nuclear physics, techniques from quantum information science are now being used to understand confinement and develop predictive capabilities for the resulting dynamical phenomena.

We have performed classical simulations of string formation and breaking in the (Abelian) Schwinger model, finding that quantum complexity varies rapidly during string breaking, and provides distinct probes complementary to classical quantities like energy density.
Importantly, nonlocal 
quantum complexity (as measured by both the MI and the NL RoM) is found along the string.
This shows that both classical and quantum correlations develop during string breaking, and suggests that the longer distance 
correlations are predominately classical.
Our nonlocal complexity computations are limited to small systems due to the resources required for subregion minimizations and stabilizer polytope considerations.
For these reasons, we have not computed higher-body nonlocal magic measures and cannot rule them out.

The quantum complexity exhibits sharp changes in the transition from flux tubes to isolated hadronic bound states, akin to phase transitions~\cite{Grieninger:2025rdi}. 
Dynamical simulations of string breaking, via wavepacket evolution~\cite{Farrell:2024fit,Zemlevskiy:2024vxt,Farrell:2025nkx} or time-dependent Hamiltonians~\cite{Farrell:2024mgu,Li:2025sgo}, would clarify mechanisms of charge extraction and confinement that are difficult to investigate with statics alone. 
Applying our techniques to non-Abelian and higher-dimensional systems will be illuminating.
We also anticipate that the quantum complexity in string breaking will be imprinted in final states of 
collisions of hadrons and nuclei.  
Analyses quantifying entanglement and magic in top-anti-top production events at the LHC are already underway~\cite{White:2024nuc,Yazgan:2025pah}.
If detected in experiment, the hierarchies of these correlations and their (non)locality may affect modeling of fragmentation and hadronization.

 \vskip 0.05in

\noindent
{\it Acknowledgements---}We would like to thank Roland Farrell, Henry Froland, Tobias Haug, Dima Kharzeev, Eliana Marroquin and
Caroline Robin for helpful discussions. 
This work was supported, in part, by U.S. Department of Energy, Office of Science, Office of Nuclear Physics, InQubator for Quantum Simulation (IQuS) under Award Number DOE (NP) Award DE-SC0020970 via the program on Quantum Horizons: QIS Research and Innovation for Nuclear Science.
S.G. was supported in part by the U.S. Department of Energy, Office of Science, Office of Nuclear Physics, Grants No. DE-FG02-97ER-41014 (UW Nuclear Theory) and in part by a Feodor Lynen Research fellowship of the Alexander von Humboldt foundation. 
This work was also supported, in part, by the Department of Physics and the College of Arts and Sciences at the University of Washington.
This work was enabled, in part, by the use of advanced computational, storage and networking infrastructure provided by the Hyak supercomputer system at the University of Washington.
This research used resources of the National Energy Research Scientific Computing Center (NERSC), a Department of Energy Office of Science User Facility using NERSC award NP-ERCAP0032083.
This research was done using services provided by the OSG Consortium \cite{osg07, osg09, https://doi.org/10.21231/906p-4d78, https://doi.org/10.21231/0kvz-ve57}, which is supported by the National Science Foundation awards \#2030508 and \#1836650.
We have made use of the ITensor library for tensor network computations~\cite{ITensor,ITensor-r0.3,Corbett:2025flm}.

\clearpage
\onecolumngrid
\appendix

\section{Analysis of lattice spacing artifacts}
\label{app:lattice_artifacts}
\begin{figure}[ht!]
    \centering
     \includegraphics[width=1.0\linewidth]{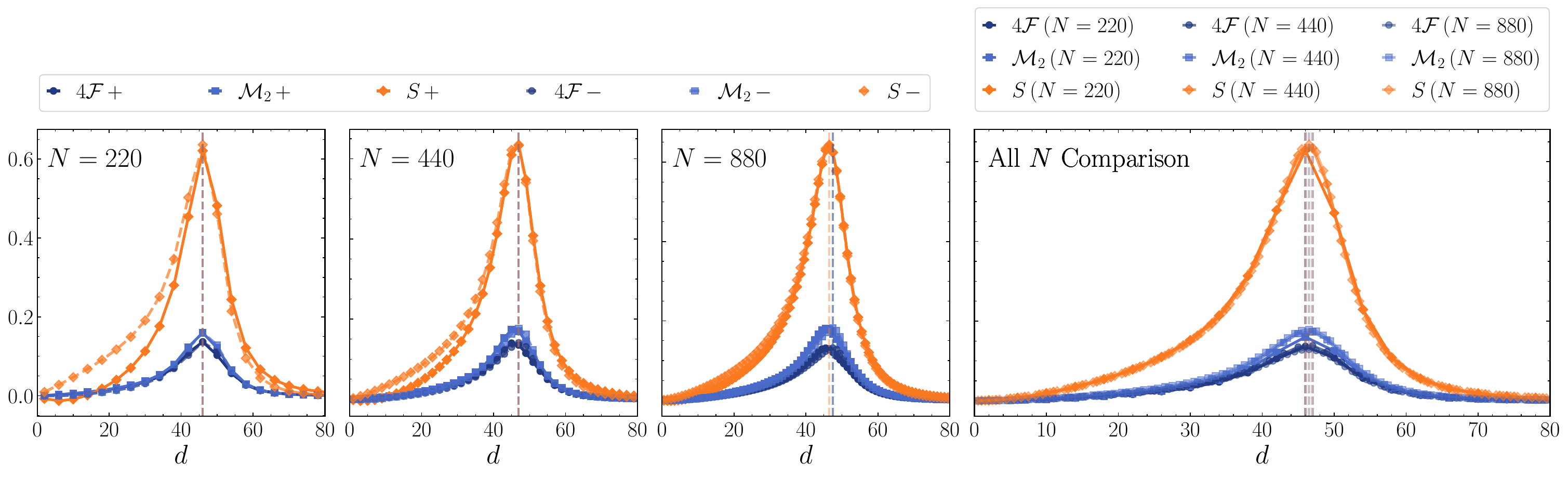}
     \caption{Bipartite measures of entanglement and quantum complexity as a function of $N,a$ for $(N,a)=(220,1), (440,1/2)$, and $(880,1/4)$.
     The antiflatness $4{\cal F}$, the upper bound to the nonlocal magic ${\cal M}_2$, and the entanglement entropy $S$ are shown for both orientations of the external field $E\text{ext}=-$ and $E_\text{ext}=+$.
     The right panel shows the average over the orientations of $E_\text{ext}$ for all $N,a$.
     The locations of the peaks for each measure are given in Table~\ref{tab:peaks}.}
     \label{fig:quantsN}
\end{figure}

\noindent
On a staggered lattice, the electric field points from antifermion to fermion sites. For our choice of parameters, odd sites correspond to fermions and even sites to antifermions. Hence, for $N/2$ even the middle link is between an antifermion and fermion site and the external electric field should point to the right (i.e., larger indices in the chain). This corresponds to our $d=a$ configuration. For $d=3a$, the opposite configuration is supported. Note that both the $d=a$ and $d=3a$ configuration belong to the same physical site. This implies that we should average over neighboring configurations to obtain physical sites, i.e., we average $d=a$ and $d=3a$, $d=5a$ and $d=7a$, etc. The averaged configurations then correspond to physical separations of $d=2a, 6a,$ etc. In the continuum, there is no distinction between fermion and antifermion sites and the results obtained for either direction of the electric field should be equal. Hence, a good proxy for the impact of lattice artifacts is to compare the results obtained for both directions of the external electric field. This is exemplified in Fig~\ref{fig:quantsN}. In the first three panels, 
we keep $L=N a$ fixed and decrease the lattice spacing from $a=1$ to $a=1/4$. For $a=1$ there is a clear difference in the entanglement entropy for the two different directions of the $E_\text{ext}$. In fact, the $S_+$ curve even turns negative at small $d$. The two curves converge in the $a\to 0$ limit. To account for the difference at finite lattice spacing, we average $S,\, \cal{F}$ and ${\cal M}_2$ over the two different directions of the electric field as shown in the fourth plot in Fig~\ref{fig:quantsN}.

\begin{table}[h]
\renewcommand{\arraystretch}{1.4}
\begin{tabularx}{\linewidth}{|c||Y|Y|Y||Y|Y|Y||Y|Y|Y|}
\hline
\rule{0pt}{10pt} & \multicolumn{9}{c|}{ Peak position}\\\hline\hline
\rule{0pt}{10pt} \multirow{2}{*}{Measure} &  \multicolumn{3}{c||}{$N=220,a=1$}  &  \multicolumn{3}{c||}{$N=440,a=1/2$} & \multicolumn{3}{c|}{$N=880,a=1/4$}  \\\cline{2-10}
& $E_\text{ext}=-$ &  $E_\text{ext}=+$ & avg. & $E_\text{ext}=-$ & $E_\text{ext}=+$ & avg. & $E_\text{ext}=-$ & $E_\text{ext}=+$ & avg. \\
\hline\hline
${\cal F}$ & 46.0 & 46.0 & 46.0  & 47.0 & 45.5 & 46.25 & 45.5 & 47.5 & 46.5 \\\hline
${\cal M}_2$ & 46.0 & 46.0 & 46.0 &  47.0 & 45.0 & 46.0 & 45.5 & 47.5 & 46.5 \\\hline
$S$ & 46.0 & 46.0 & 46.0 & 47.0 & 47.0 & 47.0 & 46.5 & 46.5 & 46.5 \\
 \hline
\end{tabularx}
\caption{The values of $d$ for which the measures (left column) reach their maximum value, for parameters in the main text and both orientations of the external field $E_\text{ext}=-$ and $E_\text{ext}=+$, and the average of the two orientations.}
\label{tab:peaks}
\end{table}

\section{OBC vs PBC}
\label{app:obc_vs_pbc}
\begin{figure}[ht!]
    \centering
     \includegraphics[width=0.75\linewidth]{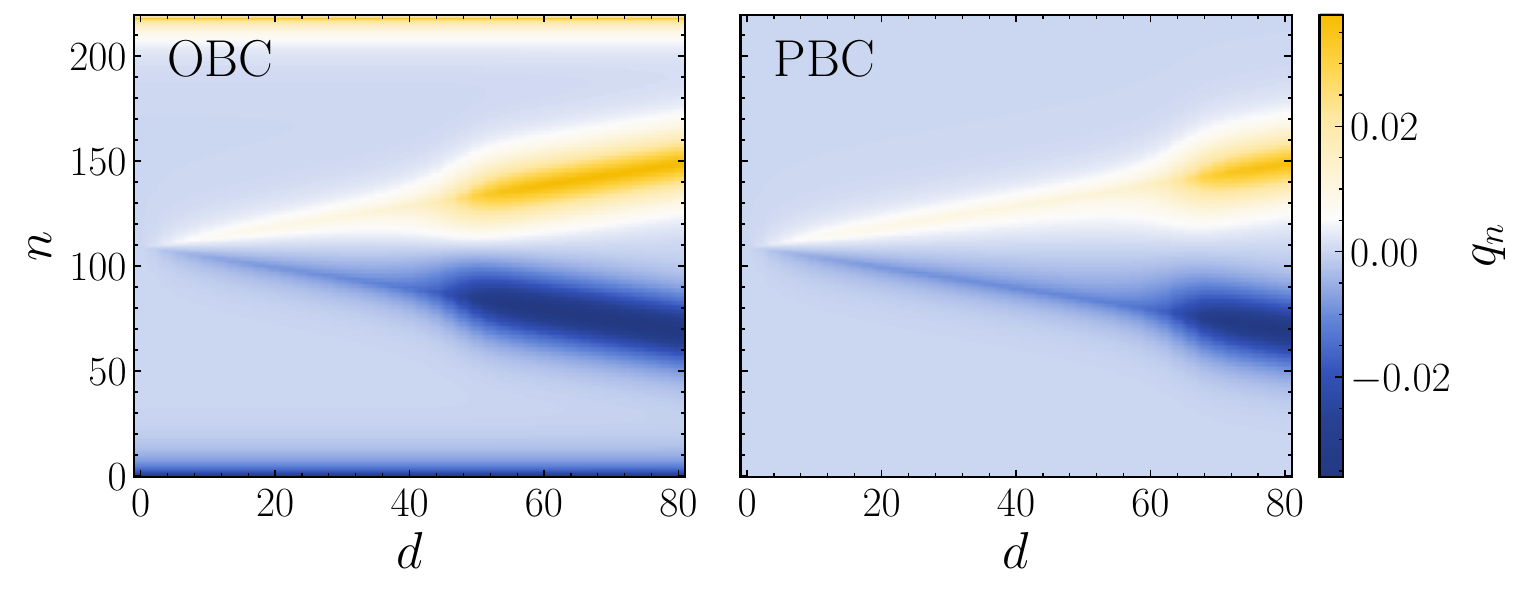}
     \caption{The charge density for OBC (left) and PBC (right) $N=220,a=1,m_\text{lat}=0.045,g=0.09$.}
     \label{fig:obc_vs_pbc}
\end{figure}

\noindent
With the parameters chosen as described in the main text, the effect of boundaries and finite volume on the physics of the simulation is exponentially suppressed.
As a result, the choice of boundary conditions should not impact the results of the computations.\footnote{OBCs are highly preferred due to the inefficiency of standard MPS in representing periodic states~\cite{PhysRevLett.93.227205}.} 
With OBCs, dynamical gauge field degrees of freedom can be eliminated from the theory, resulting in long range fermionic interactions through a linear potential. 
In PBCs, there is a single gauge field degree of freedom, the ``zero mode'', whose dynamics is not constrained by Gauss's law.
We follow the approach of Refs.~\cite{Farrell:2025nkx,Zache:2018cqq,Zache:2020qny}, where the gauge field zero mode is truncated to a single state and zero mode dynamics is not considered.
This approximation is valid for low-energy dynamics, and corrections are suppressed by $L$.
The Hamiltonian is modified from Eq.~\eqref{eq:h_obc}:
\begin{align}
    \hat{H}_\text{PBC}\,(d) \ &= \ \frac{m_\text{lat}}{2}\sum_{n=1}^{N}(-1)^n \hat{Z}_n
- \frac{g^2 a}{2}\sum_{n=1}^{N}\left\{\sum_{s=1}^{N_\text{phys}}
\left(s - \frac{s^2}{N}\right)
\left(1 - \frac{\delta_{s,N_\text{phys}}}{2}\right)
\hat{Q}_n\hat{Q}_{n+s} + \frac{2}{g}\sum_{k=1}^{N-1}\frac{N-k}{N}E_{\text{ext},n}(d)\hat{Q}_{n+k}\right\} \nonumber \\
&\quad
+ \frac{1}{4a}\sum_{n=1}^{N-1}
\left(\hat{X}_n \hat{X}_{n+1} + \hat{Y}_n \hat{Y}_{n+1}\right)
+ \frac{1}{4a}(-1)^{N_\text{phys}+1}
\left(\hat{X}_{N} \hat{X}_{1} + \hat{Y}_{N} \hat{Y}_{1}\right) \ .
\end{align}
With the lattice parameters chosen as in the main text, we observe a difference in the string breaking process between PBCs and OBCs with the same system size $N$.
As shown in Fig.~\ref{fig:obc_vs_pbc}, the string breaks at a larger separation $d$ for PBCs. 
This is attributed to image charge effects in PBCs that are absent in OBCs,\footnote{This is verified by considering a larger lattice with the same lattice spacing, where the image charges are further away and the string breaks at an $d$ closer to OBCs.} indicating that these calculations are not far into regime prescribed in the main texts.
The boundary effects, both from the lattice boundary in OBCs and image charges in PBCs, extend into the lattice volume by a length scale set by the lightest hadron mass.
As expected, the charge density of the OBC and PBC vacua determined by DMRG agree in the bulk.

\section{Charge Rearrangement During String Breaking}
\label{app:charge_rearrangement}
\noindent
During string breaking, charge is extracted from the vacuum to screen the heavy charges, so it is natural to study the rearrangement of charge. 
Considering a bipartition of the system at the center of the lattice, the ground state wavefunction may be written as in Eq.~\eqref{eq:ABQ}.
The charge rearrangement may be examined through the effect that screening has on the sector weights $p_Q$.
At $d=0$, the ground state is the lowest-lying state with total $Q=0$, so that only charge-neutral terms appear in Eq.~\eqref{eq:ABQ}.
Introducing background charges causes the weights $p_Q$ to change, as the vacuum rearranges to the new lowest-energy state.
When the background charges are far apart, effects near the bipartition are exponentially suppressed by confinement, so only the light degrees of freedom that do not participate in screening contribute to $p_Q$.
The half-lattice charge changes by one. 
In the following, $|Q^+\rangle$ denotes an external charge of $+1$, and $|q\rangle$ denotes the (reduced) state of the half-lattice. The changes for that the lowest several charge sectors experience are
\begin{eqnarray}
|Q^+\rangle\otimes |q=0\rangle &\rightarrow & |Q^+e^-\rangle \otimes |q=+1\rangle
 \ , \nonumber\\
|Q^+\rangle\otimes |q=+1\rangle &\rightarrow & |Q^+e^-\rangle \otimes |q=+2\rangle
 \ , \nonumber\\
|Q^+\rangle\otimes |q=-1\rangle &\rightarrow & |Q^+e^-\rangle \otimes |q=0\rangle
 \ ,  \label{eq:charge_transitions}
\\
 |Q^+\rangle\otimes |q=+2\rangle &\rightarrow & |Q^+e^-\rangle \otimes |q=+3\rangle
 \ , \nonumber\\
|Q^+\rangle\otimes |q=-2\rangle &\rightarrow & |Q^+e^-\rangle \otimes |q=-1\rangle
 \ , \nonumber
\end{eqnarray}
and similarly for the $|Q^-\rangle$ region.
In other words, for the half of the lattice with the $|Q^+\rangle$ external charge, ${p_{q=0}\to p_{q=+1}}$, ${p_{q=+1}\to p_{q=+2}}$, etc.

\begin{figure}[ht!]
     \centering
     \includegraphics[width=0.75\linewidth]{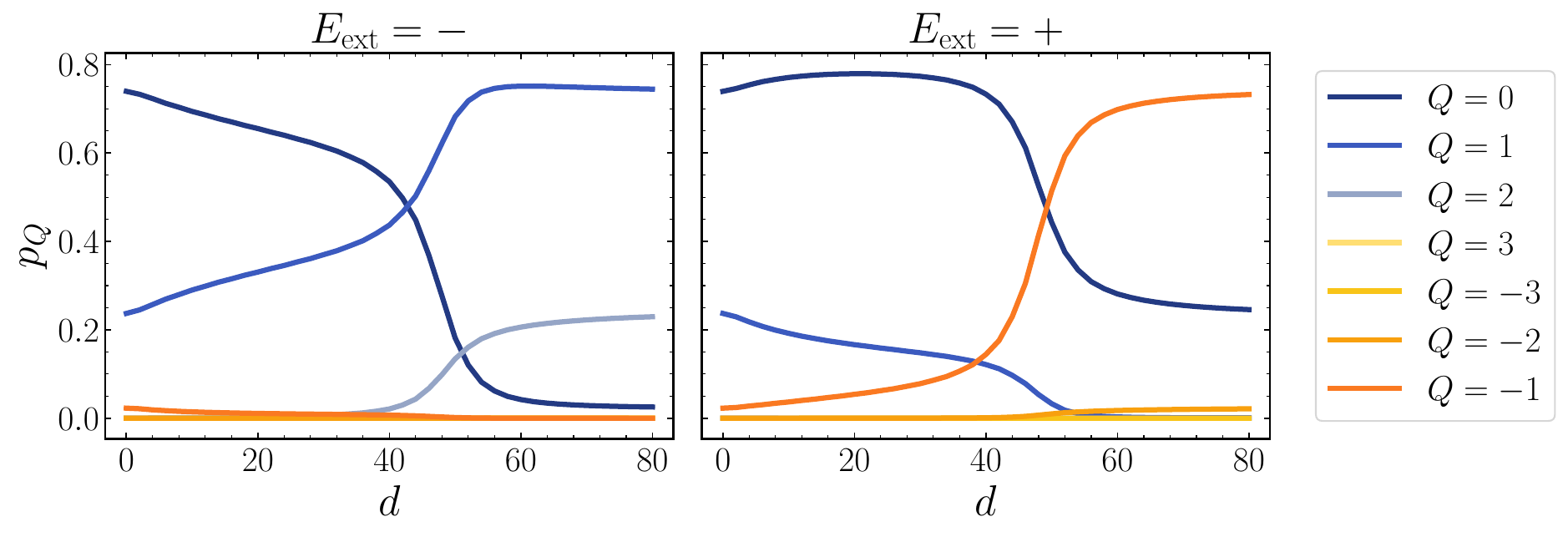}
     \caption{The charge sector weights $p_Q$ as a function of external charge separation $d$ for the two different orientations of the external field, $E_\text{ext}=-$ and $E_\text{ext}=+$. The system parameters $N=220,a=1,m_\text{lat}=0.045,g=0.09$ are used.}
     \label{fig:charge_spectrum}
\end{figure}

In our formalism, string breaking may be observed in two equivalent ways, with the external field pointing toward $n=0$, $E_\text{ext}=-$, and toward $n=N-1$, $E_\text{ext}=+$.\footnote{The configuration $E_\text{ext}=-$ corresponds to an external fermion on the left half of the lattice, and an antifermion on the right, and vice-versa for $E_\text{ext}=+$.}
Fig.~\ref{fig:charge_spectrum} shows the rearrangement of $p_Q$ for the two choices of $E_\text{ext}$.

Physical quantities should not depend on the choice of $E_\text{ext}$ orientation. 
However, we observe that the $p_Q$ transitions do not follow Eq.~\eqref{eq:charge_transitions} exactly, and behave differently for the different orientations of the electric field.
OBCs generate a charge asymmetry along the lattice from the ordering of fermions and antifermions (seen Fig.~\ref{fig:obc_vs_pbc}), created by configurations localized around the boundaries.
These boundary charge densities present a challenge for interpreting global measures of entanglement and quantum complexity, that have sensitivity to the entire lattice or half-lattices.
This effect is responsible for the differences with Eq.~\eqref{eq:charge_transitions} and the asymmetry of the left and right panels of Fig.~\ref{fig:charge_spectrum}.

Attempts to compensate for boundary effects by inserting image charges did not reduce the asymmetry between the two panels of Fig.~\ref{fig:charge_spectrum}.
Switching to PBCs as described in App.~\ref{app:obc_vs_pbc} introduces a second entangling surface, and makes Eq.~\eqref{eq:charge_transitions} superficially true due to symmetry. 
Considering a subregion of the bipartition that does not include the external boundary has similar issues. 
As seen in the left panels of Fig.~\ref{fig:quantsN} and Table~\ref{tab:peaks}, these effects do not go away with reduced lattice spacing, as naively expected. 
For this reason, it is important to consider measures of quantum complexity that are insensitive to boundary effects.

\section{Additional Classical Observables}
\label{app:classical}
\begin{figure}[ht!]
     \centering
     \includegraphics[width=0.6\linewidth]{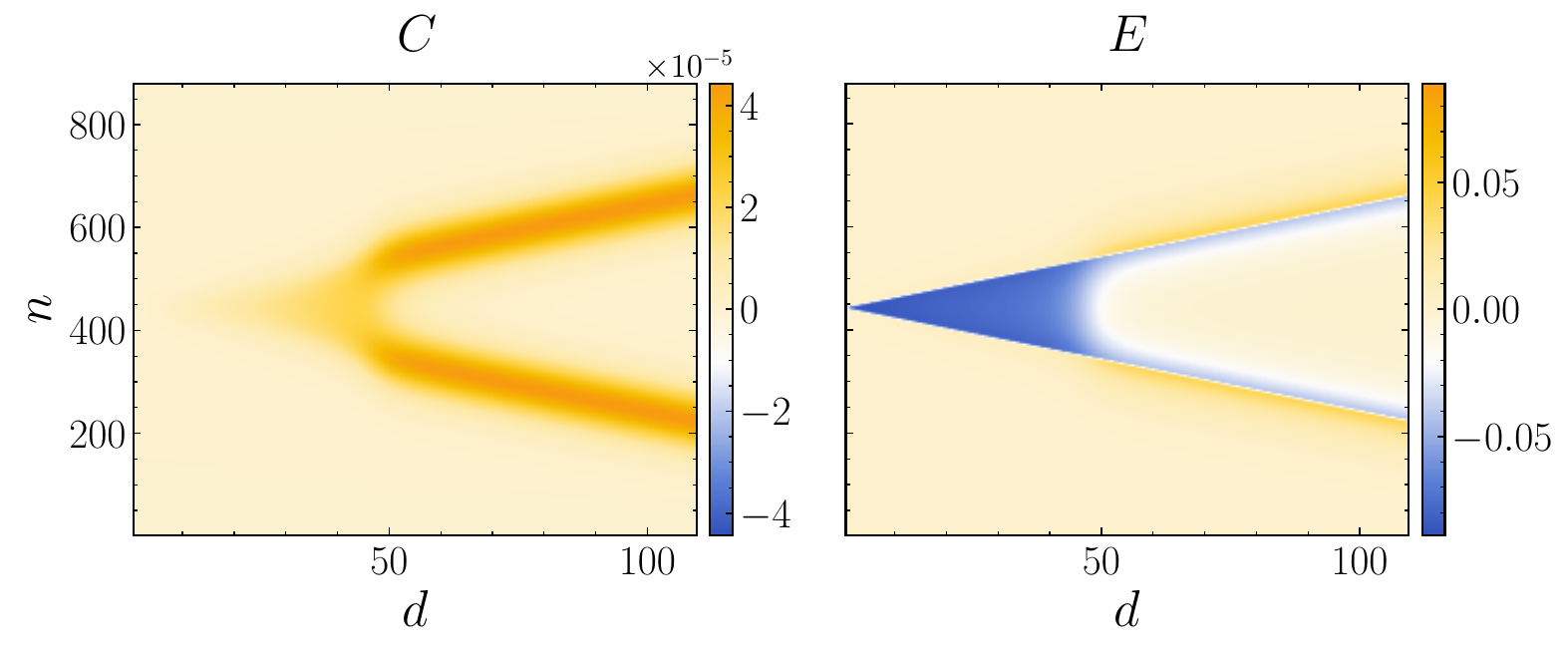}      
     \caption{The vacuum-subtracted chiral condensate (left panel) and electric field (right panel)
     obtained with simulation parameters described in the text using $N=880$ and $a=1/4$.
}
     \label{fig:cond}
\end{figure}

\noindent
Figure~\ref{fig:cond} shows the chiral condensate and the electric field as functions of position $n$ and separation of external charges $d$. The chiral condensate $C$ is defined as
\begin{align}
    C_n \ &= \ \frac{(-1)^n}{2a}\langle Z_n\rangle  \ ,
\end{align}
and the electric field $E_n$ is given in Eq.~\eqref{eq:e_field_charge}.
The relevant components of the energy-momentum tensor are given by~\cite{Grieninger:2025mbm}
\begin{align}
    \hat T^{00}_n&=\,
      \frac{1}{2a}\left( \hat K_n+\hat K_{n-1}+m(-1)^n \hat Z_n 
     \,+ \frac{a}{2}\left(\hat E_{\text{tot},n}^2+\hat E_{\text{tot},n-1}^2\right) \right),\\
         \hat T^{11}_n&=\,
      \frac{1}{2a}\left( \hat K_n+\hat K_{n-1}- \frac{a}{2}\left(\hat E_{\text{tot},n}^2+\hat E_{\text{tot},n-1}^2\right) \right),
\end{align}
Here, $\hat K_n=\frac{1}{4a} (\hat X_{n+1}\hat X_n+\hat Y_{n+1}\hat Y_n)$ is the kinetic term, $\hat E_{\text{tot},n}=\hat E_n+E_{\text{ext}, n}$ and all indices that are smaller than 1 or larger than $N$ are zero with open boundary conditions. 
The third component $\hat T^{01}_n$ vanishes in the ground state.
The energy density $\varepsilon$ and the pressure p are related to the energy-momentum tensor components by
\begin{align}
\hat\varepsilon_n &= \frac{1}{2}\left(\hat T^{00}_n -\hat  T^{11}_n \pm \sqrt{(\hat T^{00}_n + \hat T^{11}_n)^2 - 4(\hat T^{01}_n)^2}\right), \\
\hat p_n &= \hat \varepsilon - (\hat T^{00}_n - \hat T^{11}_n),
\end{align}
where the positive sign is chosen for $\hat T^{00}_n +\hat  T^{11}_n\ge0$ and the negative sign otherwise.

\begin{figure}[ht!]
     \centering
     \includegraphics[width=0.6\linewidth]{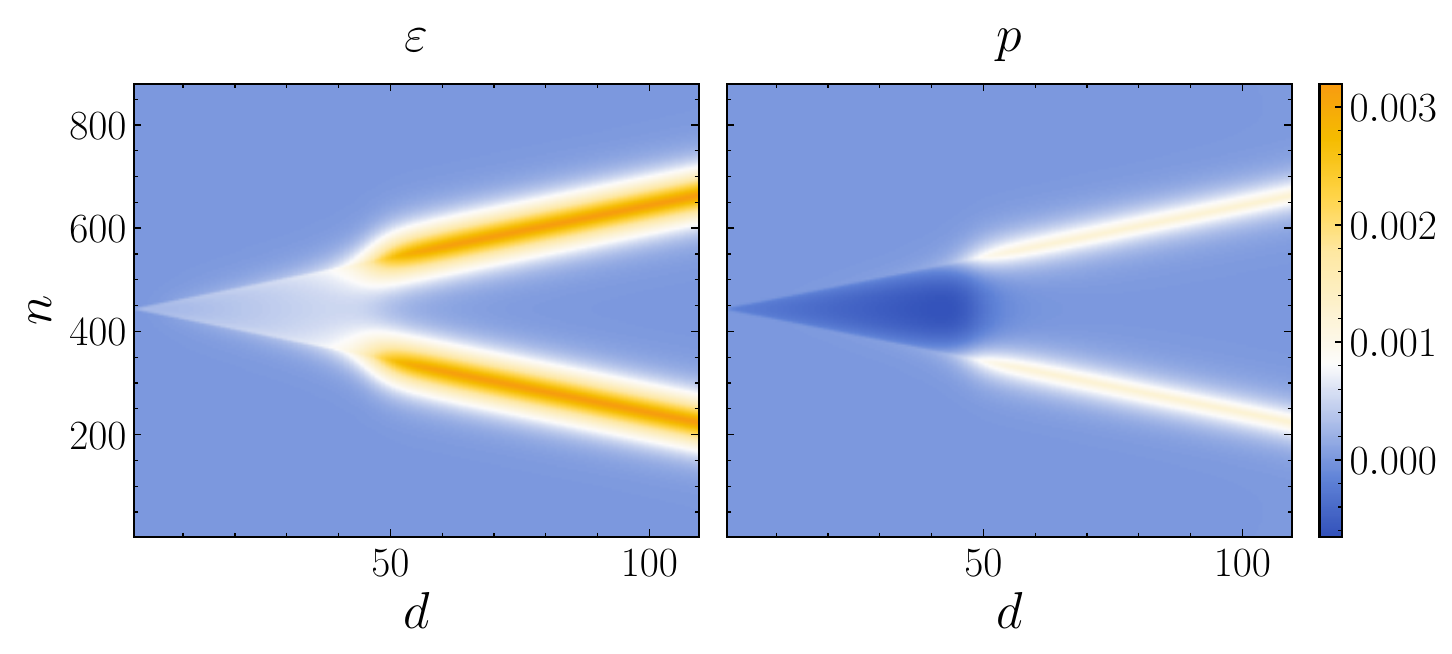}
     \caption{Components of the vacuum-subtracted energy-momentum tensor obtained with simulation parameters described in the text using $N=880$ and $a=1/4$.
     The left panel shows the energy density $\varepsilon$, while the right panel shows the pressure $p$.
     }
     \label{fig:emten}
\end{figure}

We computed the energy density and pressure, which are displayed in Fig.~\ref{fig:emten}.
The results are consistent with those given in Ref.~\cite{Grieninger:2025rdi}, but with finer resolution of the string-breaking process.  
The pressure distribution is particularly interesting.
The pressure is seen to become increasingly negative with increasing static-charge separation, and then revert rapidly to the vacuum value after string breaking.  In contrast, the pressure within the mesons rapidly increases from its vacuum  value after string breaking.

\section{Supplemental Quantum Complexity Results}
\label{app:supplemental_results}
\noindent 

\begin{figure}[ht!]
     \centering
     \includegraphics[width=0.8\linewidth]{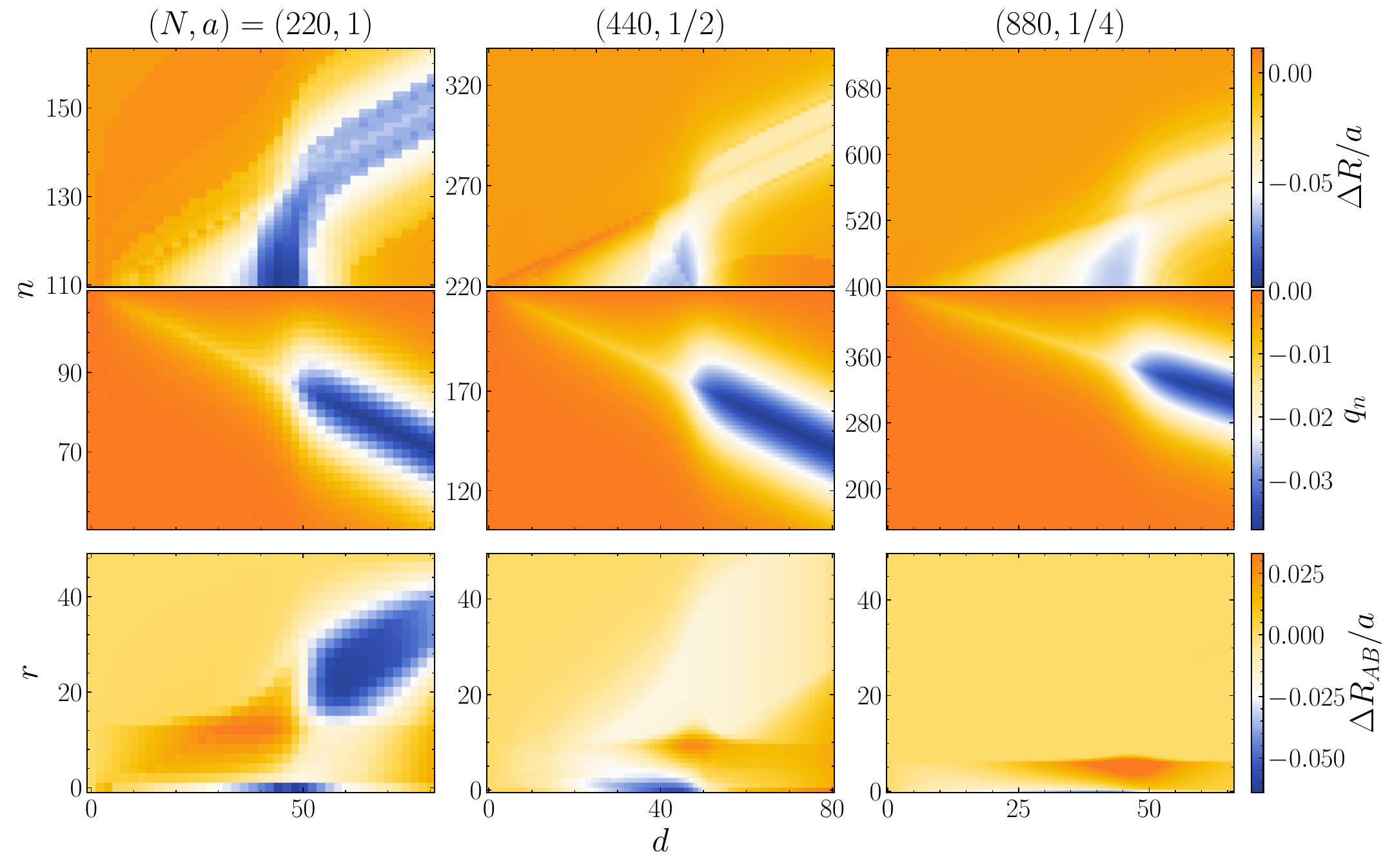}
     \caption{The dependence of the RoM on system size and lattice spacing.
     Top: The vacuum-subtracted RoM of adjacent sites, $\Delta R$ for a selection of system sizes and lattice spacings $(N,a)=(220,1)$ (left), $(440,1/2)$ (center), and $(880,1/4)$ (right). 
     The RoM is plotted in units of $a$ for half of the lattice $n\geq N/2$ as a function of external charge separation $d$.
     The system parameters from the main text are used.
     Middle: the associated charge densities for half of the lattice $n\leq N/2$.
     Bottom: the RoM of disjoint regions, $\Delta R_{AB}$, with region $A$ at the center of the lattice and region $B$ a distance $r$ away. 
     The vacuum value ($d=0$) and the value as $r\to\infty$ are subtracted.
     The RoM is plotted in units of $a$ for the same system parameters.
     }
     \label{fig:ROMadjALL}
\end{figure}

\noindent A natural quantity to consider is the behavior of magic in the continuum as $a\to0$. In principle, this could be studied by computing the RoM of a region of fixed physical volume, (i.e., $L=N_\text{phys}a = const.$).
However, the number of stabilizer states over which the RoM must be computed grows superexponentially~\cite{Aaronson:2004xuh}, and is intractable for more than 8 qubits~\cite{Hamaguchi:2023zpb}.
Instead of this, we compute the RoM for a subregion of fixed size, independent of $a$. 
While this does not probe the continuum limit of RoM in a physical region, it probes finer components of the wavefunction that cannot be seen with smaller system sizes.
Figure~\ref{fig:ROMadjALL} shows the RoM of adjacent physical sites (4 qubits) $\Delta R$ as a function of the position $n$ and separation of external charges $d$.
The plots with smaller $a$ provide a finer, ``zoomed in'' resolution view into the structure of the states. 
At $N=440,a=1/2$, additional structure is seen in $\Delta R$ between the mesons after the string has broken.
Interestingly, this structure resembles the structure in the pressure shown in Fig.~\ref{fig:emten}. 
The bottom row of Fig.~\ref{fig:ROMadjALL} shows the RoM between disjoint regions of the lattice as a function of separation (defined in Eq.~\eqref{eq:R4AB}) $\Delta R_{AB}$, scaled by $a$.
The structure initially present in the $N=220, a=1$ data is seen to vanish as $a$ is decreased.
The fixed-size RoM calculation (4 qubits) shows less information for finer $a$ because the wavefunction is spread over more lattice sites.
Instead, only quasi-local structure (small $r$) remains. 
\begin{figure}[ht!]
     \centering
     \includegraphics[width=\linewidth]{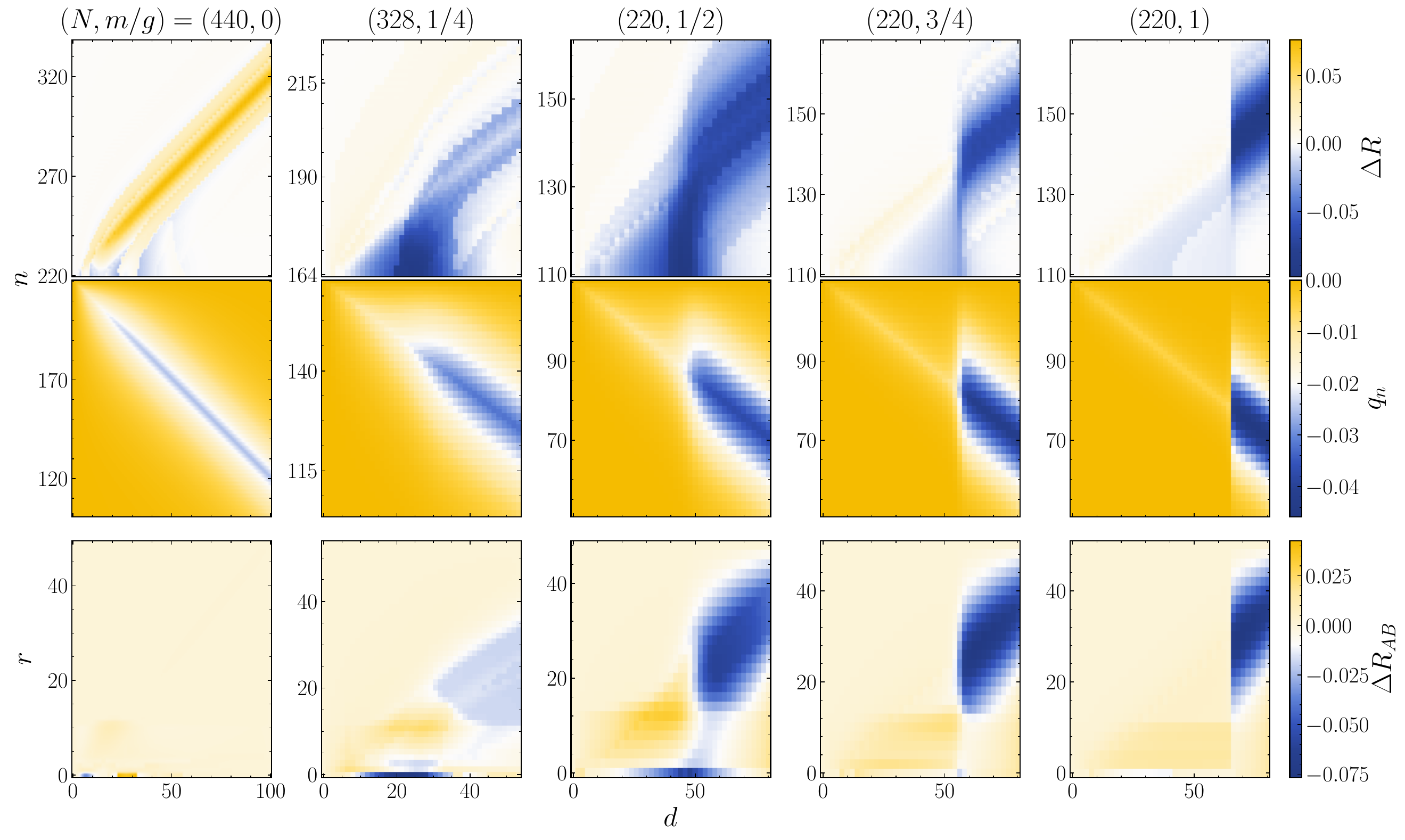}    
     \caption{The dependence of the RoM breaking on fermion mass for a selection of masses, $m/g=0$ (first column), $m/g=1/4$ (second column), $m/g=1/2$ (third column), $m/g=3/4$ (fourth column) and $m/g=1$ (fifth column).
     Top: the local vacuum-subtracted RoM $R$ as a function of $d$ for half of the lattice $n\geq N/2$ with $a=1$.
     Middle: the associated charge densities for half of the lattice $n\leq N/2$ for the same parameters.
     Bottom: the RoM of disjoint regions, $\Delta R_{AB}$, with region $A$ at the center of the lattice and region $B$ a distance $r$ away.
     The vacuum value ($d=0$) and the value as $r\to\infty$ are subtracted.
     The RoM is plotted in units of $a$ for the same system parameters.
     }
     \label{fig:m_g_scan}
\end{figure}

Figure~\ref{fig:m_g_scan} details the behavior of the quantum complexity as $m/g$ is varied.
We consider several values of $m$ holding $g$ fixed as in the main text, and expand the lattice to account for the increased correlation length with smaller mass.
The top and middle rows display the RoM between adjacent sites, $\Delta R$, and the charge density $q_n$ as a function of position $n$ and separation $d$. 
The transition from strings to mesons becomes increasingly sharper with larger $m/g$, which is seen in both $\Delta R$ and $q_n$.
The string also appears to break at a larger separation $d$ for larger $m/g$.
This is an expected result of fermions being more ``expensive'' to excite from the vacuum to screen the external charges.
The bottom row, as in Fig.~\ref{fig:ROMadjALL}, shows the RoM between disjoint $\Delta R_{AB}$ subsystems on the lattice as a function of the distance between the regions $r$.
Region $A$ is at the center of the lattice, and region $B$ is distance $r$ away.
The vacuum-subtracted $\Delta R_{AB}$ only shows structure at small $r$ for the massless case, consistent with the system only having short-range quantum correlations before string breaking. 
At intermediate $m/g$, the correlations before string breaking are longest-range, and then die off again at $m/g=1$.
The propagation of the bound hadrons is seen in all parameters besides the massless case.

\section{Computational Methods}
\label{app:mps_methods}
\noindent
The reduced density matrix $\hat{\rho}_{AB}$ is required for the nonlocal measures of complexity such as MI and RoM, where regions $A$, $B$ may be separated on the lattice.
To avoid exponentially-scaling resource requirements for creating a long-distance reduced density matrix, we apply a swap network to the MPS to move region $B$ next to region $A$, and then trace the regions outside $A$, $B$. 
This introduces long-range entanglement in the system and has the same cost as exactly contracting the region between A and B.
We approximate this operation by introducing a bond dimension and cutoff for the swap network operation.

We find ground states and low-lying excited states in MPS with DMRG. 
We use a bond dimension of 400, a cutoff of $10^{-12}$ and 40 sweeps. 
The convergence of our MPS computations, both for DMRG and for computing nonlocal reduced states is verified by examining the results as the precision is increased.

The RoM is calculated using the approach of Ref.~\cite{Hamaguchi:2023zpb}.
The minimization problem of Eq.~\eqref{eq:RoMdef} is reframed as a sparse linear programming problem,
\begin{align}
    R(\hat{\rho}) = \min_{\bm x} \left\{
    \ ||\bm x||_1 \ \ \  \bigg\rvert \ \ \  \ \hat{A} \bm x = \bm b\right\} \ .
    \label{eq:rom_axb}
\end{align}
Here the matrix $\hat{A}$ is defined as $ \hat{A}_{ij}=\text{Tr}(\hat{P}_i \hat{\rho}_{s_j})$, and $\bm b$ is the expansion of the state $\hat{\rho}$ in the Pauli basis: $b_i=\text{Tr}(\hat{P}_i\hat{\rho})$.
In practice, this may be readily solved in the following form:
\begin{align}
    \underset{\bm u} {\rm minimize} \ \sum_i u_i \ , \ \text{s.t.} \ \left( \hat{A} \quad - \hat{A}\right)\bm u=\bm b \ , \  u_i \geq 0 \ ,
\end{align}
A further simplification developed in Ref.~\cite{Hamaguchi:2023zpb} is that only a small fraction of stabilizer states have $x_i>0$, and these are correlated with having large values of the overlap $|2^{n_Q}\text{Tr}\left(\hat{\rho}_{s_i}\hat{\rho}\right)|$.
This enables further sparsification of the problem, only retaining a fraction $K$ of the states $\hat{\rho}_{s_i}$.
In this work, $K=0.05$ of states with the largest overlaps are kept for the optimization problem.

The NL RoM is calculated as the minimum of 1000 optimizations as in Eq.~\eqref{eq:NLRoMdef}, each with randomly chosen initial conditions.

\bibliography{bibi}

\end{document}